\documentclass[12pt,a4paper]{article}
\usepackage{amsfonts}
\usepackage{amsmath}
\usepackage{amssymb}
\usepackage{latexsym}

\numberwithin{equation}{section}

\providecommand{\abs}[1]{\lvert#1\rvert}

\newcommand{\BbbR}{\mathbb{R}}
\newcommand{\BbbZ}{\mathbb{Z}}

\newcommand{\RPthree}{\mathbb{RP}^3}
\newcommand{\RPtwo}{\mathbb{RP}^2}

\newcommand{\rmd}{\mathrm{d}}
\newcommand{\rme}{\mathrm{e}}

\newcommand{\myell}{\ell}

\newcommand{\Mcft}{M_{\mathrm{CFT}}}

\newcommand{\rmin}{r_{\mathrm{min}}}
\newcommand{\rmax}{r_{\mathrm{max}}}
\newcommand{\Lren}{L_{\mathrm{ren}}}

\newcommand{\Gwide}{{\widetilde{G}}}

\newcommand{\lanln}[1]{$\langle$\texttt{arXiv:#1}$\rangle$}

\newcommand{\hdbtz}{HD-BTZ}

\newtheorem{lemma}{Lemma}[section]

\newcommand{\myproof}{\emph{Proof\/}. }

\title{Einstein black holes, 
free scalars 
and 
\\
AdS/CFT correspondence}

\author{Jorma Louko\thanks{jorma.louko@nottingham.ac.uk} 
\and Jacek Wi\'sniewski\thanks{jacek.wisniewski@nottingham.ac.uk~. 
Address after 1 September 
2004: The Boston Consulting Group,
Warsaw, Poland; jacwisn@yahoo.com~.}
\\
\noalign{\vspace{3ex}}
{\it School of Mathematical Sciences,
University of Nottingham,}\\
{\it Nottingham NG7 2RD, UK}
\\
\noalign{\vspace{3ex}}\\
\small{(Revised August 2004)}
\\
\noalign{\vspace{2ex}}
\small{\lanln{hep-th/0406140}}
\\
\noalign{\vspace{2ex}}
\small{Published in {\it Phys.\ Rev.\ \rm D \bf 70}, 084024 (2004)}
}

\date{}

\begin{document}

\maketitle

\begin{abstract}
We investigate AdS/CFT correspondence for two families of Einstein black
holes in $d\ge4$ dimensions, modelling the boundary CFT by a free conformal
scalar field and evaluating the boundary two-point function in the bulk
geodesic approximation. For the $d\ge4$ counterpart of the nonrotating BTZ
hole and for its $\BbbZ_2$ quotient, the boundary state is
thermal in the expected sense, and its stress-energy reflects the properties
of the bulk geometry and suggests a novel definition for the mass of the
hole.  For the generalised Schwarzschild-AdS hole with a flat horizon of 
topology~$\BbbR^{d-2}$, the boundary stress-energy has a thermal 
form with energy density proportional to the hole ADM mass, 
but stress-energy corrections from compactified horizon
dimensions cannot be consistently included at least for $d=5$. 
\end{abstract}

\newpage 

\section{Introduction}

In the study of AdS/CFT correspondence~\cite{malda-conj,wittensupp},
much attention has focussed on situations where the gravitational side
is a black hole solution to an appropriate (super)gravity theory. The
dual conformal field theory (CFT) 
is then expected to contain not only
information about the causal structure of the black hole but also
information about its quantum properties, in particular the Hawking
temperature. While it is difficult to address these questions strictly
within the supergravity/CFT pairs in which the evidence for duality is
the strongest~\cite{malda-conj,aharonyetal,BMN}, simplified models
that aim to capture aspects of the correspondence have been analysed
in various contexts
\cite{MAST,%
keskivakkuri,danielsson-etal1,danielsson-etal2,danielsson-etal3,%
horo-itz,BalaBoer-etal,%
louko-marolf,Bala-Ross,lou-ma-ross,louko-polanica,malda-eternal,%
sonnen-rev,gregory-ross,husain-loop,CNO,%
birmingham-quasi1,birmingham-sachs-solo,birmingham-sachs-relax,%
krasnov-holography,krasnov-acont,krasnov-holofact,%
KrausOoShen,LeviRoss,Fid-etal,PorratiRabadan,Kaplan,bala-levi}. 
In certain situations a Wilson loop on
the conformal boundary can be evaluated semiclassically from extremal 
world sheets in the bulk, relating critical phenomena in the boundary
theory to the classical properties of the bulk black hole 
\cite{sonnen-rev,gregory-ross,husain-loop,CNO}. 
In other situations it has been argued
that a boundary two-point function can be evaluated
semiclassically from bulk geodesics
\cite{Bala-Ross,lou-ma-ross,%
KrausOoShen,LeviRoss,Fid-etal,PorratiRabadan,Kaplan}, 
enabling one to relate the bulk geometry to properties of the
boundary state that are recoverable from the two-point function. 

The purpose of this paper is to examine the boundary two-point
function evaluated semiclassically from bulk geodesics in two families
of locally asymptotically anti-de~Sitter (AdS) Einstein black holes in
$d\ge4$ dimensions. One aim is to examine how consistently the
boundary theory can be modelled by a free conformal scalar field:
While a free boundary field has led to consistent results when the
bulk is a black hole quotient of AdS$_3$~\cite{lou-ma-ross}, it has
been argued that for more general black holes a better boundary model
may be a high conformal weight field~\cite{Fid-etal}. A second aim is
to examine quotients under discrete isometry groups. From the bulk
point of view, such a quotient may have a reduced number of
globally-defined isometries, and in some cases a time translation
isometry may be broken in a way that cannot be detected from just the
geometry of an asymptotic
region~\cite{louko-marolf,lou-ma-ross,MW-geondata,giulini,FriedSchWi}. From
the boundary point of view the quotient introduces new geodesics that
may contribute to the boundary two-point function, and these geodesics
may probe deep bulk regions, in some cases even behind
horizons~\cite{lou-ma-ross}. It becomes thus important to understand
in what circumstances such `long' geodesics can, or indeed must, be
included.

We begin with quotients of AdS${}_d$ with
$d\ge4$ under a group generated by a single boost-like
isometry
\cite{aminneborg,holst,banados,gomberoff}. 
These spacetimes are generalisations of the nonrotating
$(2+1)$-dimensional Ba\~nados-Teitelboim-Zanelli (BTZ) hole
\cite{teitelboim,zanelli}, 
and we refer to them as the higher-dimensional BTZ holes (\hdbtz{}
holes). They are by construction locally AdS and locally
asymptotically AdS. They are black holes, but their exterior region
does not have a global timelike Killing vector, the horizon is not
stationary, and the infinity is not globally asymptotically
AdS\null. The conformal infinity is connected and has the (conformal)
metric $\mathrm{dS}_{d-2} \times S^1$, where the circumference of the
$S^1$ is determined by the magnitude of the boost.

For a free conformal scalar field, 
the global vacuum on the conformal boundary of AdS${}_d$
induces a state on the
\hdbtz{} boundary
$\mathrm{dS}_{d-2} \times S^1$. This state turns out to be the Euclidean
vacuum
\cite{chernikov-tagirov,tagirov,bousso} 
for each Fourier mode on the circle, and we refer to it as the
Euclidean vacuum. We show that a semiclassical evaluation by bulk geodesics
reproduces the Euclidean vacuum Green's function, and including
the `long' geodesics is required for this agreement. The vacuum is thermal
in the sense of the de Sitter temperature. We compute the difference
in the stress-energy tensors in the Euclidean vacua on $\mathrm{dS}_{d-2}
\times S^1$ and $\mathrm{dS}_{d-2} \times \BbbR$, showing that this
difference decays exponentially in the size of the circle. The
stress-energy also suggests a novel definition for the mass of the
\hdbtz{} hole via AdS/CFT: As the hole is 
not asymptotically stationary, an Arnowitt-Deser-Misner 
(ADM) mass is not available. 

We then repeat the analysis for a $\BbbZ_2$ quotient of the
\hdbtz{} hole, analogous to the $\RPthree$ and $\RPtwo$
geons constructed from respectively the Schwarzschild hole
\cite{MW-geondata,giulini,FriedSchWi} and the nonrotating BTZ
hole~\cite{louko-marolf}. The reduced isometry group now singles out a 
foliation of the boundary $(\mathrm{dS}_{d-2} \times
S^1)/\BbbZ_2$ by spacelike surfaces. We find that the 
stress-energy contributions from this $\BbbZ_2$ quotient vanish at
late and early times in the distinguished foliation but dominate
at intermediate times, and in the geodesic approximation the dominant
stress-energy contribution at intermediate times comes from the
`long' geodesics created by the $\BbbZ_2$ identification.  We also discuss
briefly generalisations of the \hdbtz{} holes in which the isometry
generator is not a pure boost.

Next, we turn to spacetimes obtained from the Schwarzschild-AdS family
in $d\ge4$ dimensions by replacing the round $S^{d-2}$ by flat
$\BbbR^{d-2}$~\cite{GibbWilt,lemos1,lemos2,lemos-zanchin,birmingham-topol}.
These generalised Schwarzschild-AdS spacetimes are locally
asymptotically AdS but not of locally constant curvature. The
conformal boundary consists of two copies of $(d-1)$-dimensional
Minkowski space. We evaluate the Green's function on a single boundary
component by the geodesic approximation to quadratic order in the
coordinate separation, choosing a subtraction procedure compatible
with the free scalar field Hadamard form. The stress-energy calculated
from the boundary Green's function turns out to be that of a finite
temperature free conformal field, the temperature agreeing with the
black hole Hawking temperature up to a dimension-dependent numerical
factor, and the energy density being proportional to the ADM mass of
the hole. At this level, the geodesic method is thus consistent with
modelling the boundary theory by a free conformal field even when the
spacetime is not locally AdS\null.

We then replace the flat $\BbbR^{d-2}$ by flat $\BbbR^{d-3} \times
S^1$, so that the conformal boundary consists of
two copies of $(d-1)$-dimensional Minkowski space with one periodic
spatial dimension. Specialising to $d=5$, we evaluate the leading
periodicity correction to the quadratic terms in the boundary Green's
function in the limit of large period, finding that this correction does
not satisfy the Klein-Gordon equation. Including 
the `long' geodesics is thus not consistent with
modelling the boundary theory by a free scalar field. We find a similar
conclusion for a geon-type variant whose conformal infinity consists
of a single copy of
four-dimensional Minkowski space with one periodic spatial dimension.
In both cases the `long' geodesics probe the bulk in regions
where the local geometry deviates significantly from~AdS${}_d$. 

We use metric signature $(-++\cdots)$. The quotient spacetimes of
AdS${}_{d}$ are analysed in section \ref{sec:hBTZ} and the generalised
Schwarzschild-AdS spacetimes in 
section~\ref{sec:schw-flat}. Section \ref{sec:discussion} presents
concluding remarks. Certain technical issues are deferred
to four appendices.

\section{Higher-dimensional BTZ holes and their generalisations}
\label{sec:hBTZ}

In this section we consider quotient spacetimes of AdS${}_{d}$ with $d
\ge 4$. Subsection \ref{subsec:BTZquotient} reviews the \hdbtz{}
construction
\cite{aminneborg,holst,banados,gomberoff}. 
The boundary state is analysed in subsections \ref{subsec:euclidean},
\ref{bulkBTZ} and~\ref{subsec:boundary-stress}. Subsections
\ref{subsec:hdbtz-geon} and \ref{subsec:rotHDBTZ} discuss
respectively the further $\BbbZ_2$ quotient and inclusion of
rotation.

\subsection{\hdbtz{} quotients in the bulk and on the boundary}
\label{subsec:BTZquotient}

Recall that AdS${}_d$, $d\ge2$, can be defined as the hyperboloid
\begin{equation}
-\myell^{2} 
= 
-{(X^{0})}^{2}+{(X^{1})}^{2}+\cdots+{(X^{d-1})}^{2}-{(X^{d})}^{2}
\label{eq:hyperboloid}
\end{equation} 
in $\BbbR^{d-1,2}$ with global coordinates $(X^{a})$ 
and the metric 
\begin{equation}
\rmd s^{2}=-{(\rmd X^{0})}^{2}+{(\rmd X^{1})}^{2}+\cdots+
{(\rmd X^{d-1})}^{2}-{(\rmd X^{d})}^{2} 
\ \ . 
\label{flat}
\end{equation}
The positive parameter $\myell$ sets the
curvature scale. From now on we take $d\ge4$. 

The \hdbtz{} hole is the quotient of the
region $X^{d} > \abs{X^{d-1}}$ under the group
generated by $\exp( 2\pi \lambda_+ \xi )$, 
where $\xi$ is the Killing
vector
\begin{equation}
\xi := 
X^{d-1}
\frac{\partial}{\partial{X^{d}}} 
+ X^{d} 
\frac{\partial}{\partial{X^{d-1}}}
\label{eq:xi-def}
\end{equation}
and $\lambda_+$ is a positive parameter. 
The subregion that gives the hole exterior is 
$-(X^{0})^{2}+(X^{1})^{2}+\cdots+(X^{d-2})^{2}>0$, $X^{d}>0$. 
A~convenient parametrisation of this 
subregion for our purposes is 
\begin{subequations}
\label{eq:exterior-para}
\begin{eqnarray}
X^{0}&=&\myell \sqrt{ \rho^{2}-1}
\, 
\sinh\! \left( T \right)
\ \ ,
\label{X0} 
\\
X^{i}&=&
\myell \sqrt{ \rho^{2}-1}
\, 
\cosh\!\left( T \right)
\hat{x}^{i}
\ \ ,
\label{Xi} 
\\ 
X^{d-1}&=& \myell \rho
\sinh\!\left(\lambda_{+}\varphi\right)
\ \ ,
\label{Xd-1} 
\\ 
X^{d}&=& \myell \rho
\cosh\!\left(\lambda_{+}\varphi\right)
\ \ ,
\label{Xd}
\end{eqnarray}
\end{subequations}
where $\rho>1$, $i=1,\ldots,d-2$ and 
$\hat{x}$ is a $(d-2)$-dimensional unit
vector. 
The metric in the parametrisation
(\ref{eq:exterior-para}) 
reads 
\begin{subequations}
\label{eq:BTZ-fullset}
\begin{equation}
\rmd s^{2}=
\myell^2 \left[ 
\frac{\rmd\rho^2}{\rho^2-1} 
+ (\rho^2-1) \rmd s^2_{\mathrm{dS}}
+ \lambda_+^2 \rho^2 \rmd\varphi^2
\right] 
\ \ ,
\label{BTZ}
\end{equation}
where 
\begin{equation}
\rmd s^2_{\mathrm{dS}}
:= 
-\rmd T^{2}
+
\cosh^{2}T \, \rmd\Omega^2_{d-3}
\label{eq:dSglobalmetric}
\end{equation}
\end{subequations}
and $\rmd\Omega_{d-3}^2$
is the metric on unit~$S^{d-3}$. $\rmd s^2_{\mathrm{dS}}$ 
(\ref{eq:dSglobalmetric})
is recognised as the global metric on 
$(d-2)$-dimensional
de~Sitter space of unit curvature radius. 
As $\lambda_+ \xi = \partial_\varphi$, 
the \hdbtz{} exterior is (\ref{eq:BTZ-fullset}) 
with the identification 
$(T,\rho,{\hat x},\varphi) \sim (T,\rho,{\hat x},\varphi+2\pi)$. 
The past and
future horizons are located at $\rho=1$. 

The exterior is connected and has spatial topology $S^{d-3}\times
S^1$. It has no globally-defined timelike Killing vectors,
and the area of the event horizon is increasing in time. Further
discussion of the global properties can be found
in~\cite{aminneborg,holst}. 

The metric on hypersurfaces of constant 
$\rho$ satisfies 
\begin{equation}
\left. 
\rmd s^{2}
\right|_{\rho=\mathrm{const}}
\sim 
\myell^2 \rho^2 
\rmd s^{2}_{\mathrm{CFT}}
\ \ , 
\ \ 
\rho\to\infty
\ \ , 
\end{equation}
where 
\begin{equation}
\rmd s^{2}_{\mathrm{CFT}}
:=
\rmd s^2_{\mathrm{dS}}
+
\lambda_{+}^{2}\rmd\varphi^{2}
\ \ . 
\label{CFTmetric}
\end{equation}
We adopt $\rmd s^{2}_{\mathrm{CFT}}$ (\ref{CFTmetric}) 
as the 
representative of the conformal equivalence 
class of metrics at the infinity. 
This boundary at the infinity thus the product of $(d-2)$-dimensional
de~Sitter space of unit curvature radius and a spacelike circle
of circumference $2\pi\lambda_+$.

Suppose for the moment that $\varphi$ were not 
periodic. In the coordinates $(\tau,\theta,\hat{x})$
defined by 
\begin{subequations}
\label{eq:bctransf}
\begin{align}
\rme^{2\lambda_+\varphi} 
&= 
\frac{\cos \tau - \cos\theta}{\cos \tau + \cos\theta}
\ \ , 
\\
\tanh T 
&=
\frac{\sin \tau}{\sin\theta}
\ \ , 
\end{align}
\end{subequations}
the domain that corresponds to (\ref{CFTmetric}) would be the
diamond\linebreak 
$\mathcal{D} := \left\{(\tau,\theta,\hat{x}) \bigm|
\abs{\pi/2 - \theta}  < \pi/2 - \abs{\tau} \right\}$, 
and the metric would read 
\begin{equation}
\rmd s^{2}_{\mathrm{CFT}} 
= 
\frac{1}{(\sin^2\!\theta - \sin^2\!\tau)}
\, 
\rmd
s_{\mathrm{ESU}}^2
\ \ , 
\label{CFT-ESU}
\end{equation}
where 
\begin{equation}
\rmd s_{\mathrm{ESU}}^2 
= 
-\rmd \tau^2 + \rmd \theta^2 + \sin^2\!\theta \, \rmd\Omega^2_{d-3}
\ \ . 
\label{eq:ESU} 
\end{equation}
The metric (\ref{eq:ESU}) is recognised as
the Einstein static universe metric on the conformal boundary of
AdS${}_d$~\cite{horo-maro}. 
On making $\varphi$ periodic, equations (\ref{CFTmetric})--(\ref{eq:ESU})
therefore show how the conformal boundary of the \hdbtz{} hole
is obtained as the quotient of the diamond $\mathcal{D}$ on the conformal
boundary of AdS${}_d$. We shall use this observation to characterise a
vacuum state in subsection~\ref{subsec:euclidean}.

\subsection{Euclidean vacuum on the boundary}
\label{subsec:euclidean} 

We consider a free conformal scalar field $\phi$ 
in the metric~(\ref{CFTmetric}).
We begin with nonperiodic $\varphi$ and make $\varphi$
periodic at the end of the subsection. 

As the Ricci scalar equals $(d-2)(d-3)$, 
the wave equation reads 
\begin{equation}
-\nabla^{\mu}\nabla_{\mu}\phi+\frac{(d-3)^{2}}{4}\phi=0
\ \ . 
\label{waveeq}
\end{equation}
Separating the $\varphi$-dependence as $\rme^{im\varphi}$,
$m\in\BbbR$, gives on dS${}_{d-2}$ a wave equation with the effective
mass squared term $\tfrac14 {(d-3)^{2}} + {(m/\lambda_+)}^2$. We
denote by Euclidean vacuum the vacuum whose positive frequency mode
functions reduce to the dS${}_{d-2}$ Euclidean vacuum 
positive frequency mode functions for each $m$
\cite{chernikov-tagirov,tagirov,bousso}. A normalised set of such mode
functions is 
\begin{subequations}
\label{eq:evac-modes-full}
\begin{equation}
\phi_{mnk} := 
\frac{1}{2\sqrt{2\lambda_+}}
\, 
\rme^{im\varphi}
{(\cosh T)}^{(3 - d)/2}
f_{mn}\bigl(- \tanh T \bigr) 
\, Y_{nk}
\ \ , 
\label{eq:evac-modes}
\end{equation}
where $Y_{nk}$ are the spherical harmonics on unit
$S^{d-3}$~\cite{Bate,Vil}, the index $n$ ranges over non-negative
integers, the eigenvalue of the scalar Laplacian on $S^{d-3}$ is
$-n(n+d-4)$, the index $k$ labels the degeneracy for each~$n$, and
\begin{equation}
f_{mn}(x) := 
\rme^{\frac12 \pi(i\nu - m/\lambda_+)}
\sqrt{\frac{\Gamma\bigl(\nu - im/\lambda_+ +1\bigr)}
{\Gamma\bigl(\nu + im/\lambda_+ +1\bigr)}}
\left[ 
{\mathrm{P}}_\nu^{im/\lambda_+}(x) 
+ \frac{2i}{\pi} 
{\mathrm{Q}}_\nu^{im/\lambda_+}(x) 
\right] 
\ \ , 
\label{eq:legendre-comb}
\end{equation}
\end{subequations}
where ${\mathrm{P}}_\nu^{im/\lambda_+}$ and
${\mathrm{Q}}_\nu^{im/\lambda_+}$ are the associated Legendre
functions on the cut \cite{gradstein} and $\nu = n + \frac12(d-5)$. The
linear combination of ${\mathrm{P}}_\nu^{im/\lambda_+}$ and
${\mathrm{Q}}_\nu^{im/\lambda_+}$ in (\ref{eq:legendre-comb}) is determined
by the analytic continuation properties \cite{bousso,laf-euclidean}, as
can be verified using the formulas in \cite{magnusetal}, p.~168. 

By (\ref{CFT-ESU}), the functions
$\psi_{mnk} := {(\sin^2\!\theta - \sin^2\!\tau)}^{(3-d)/4} \phi_{mnk}$
solve the conformal scalar field equation in the diamond $\mathcal{D}$ in
the Einstein static universe~(\ref{eq:ESU}). On analytic continuation 
outside~$\mathcal{D}$, it can be verified (\cite{magnusetal},
p.~168) that $\psi_{mnk}$ are bounded in the lower half-plane in
complexified~$\tau$ and hence purely positive frequency with respect to
the Killing vector~$\partial_\tau$. This
means that our Euclidean vacuum is the vacuum induced from the
global vacuum on the conformal boundary of~AdS${}_d$. 

To evaluate the Euclidean vacuum Green's function, we recall that the
Euclidean vacuum Green's function $G_{\mathrm{dS}}^{(m)}$ on dS${}_{d-2}$
for a scalar field with mass squared $\tfrac14 {(d-3)^{2}} +
{(m/\lambda_+)}^2$ reads \cite{bousso}
\begin{equation}
G_{\mathrm{dS}}^{(m)}
(T,\hat{x};T',\hat{x}')
=
\frac{
\Gamma\bigl( \frac{d-3}{2} + i \frac{m}{\lambda_+}\bigr) 
\Gamma\bigl( \frac{d-3}{2} - i \frac{m}{\lambda_+}\bigr)}
{2 {(2\pi)}^{(d-2)/2}}
\, 
{(1-Z^{2})}^{(4-d)/4}
\, 
{\mathrm{P}}^{(4-d)/2}_{-\frac{1}{2}+i m/\lambda_{+}}(-Z)
\ \ ,
\label{eq:GdS}
\end{equation}
where the coordinates on 
dS${}_{d-2}$ are as in the first two terms in~(\ref{CFTmetric}), 
\begin{equation}
Z(T,\hat{x};T',\hat{x}')
:=
\cosh(T) \cosh(T') 
(\hat{x} \cdot \hat{x}') 
- \sinh(T) \sinh(T') 
\ \ , 
\label{eq:Zdef}
\end{equation}
$\hat{x}\cdot\hat{x}' :=
\sum_{i}x^{i}{\vphantom{(}x'\vphantom{)}}^{i}$, and we have written
the hypergeometric function of \cite{bousso} in terms of the
associated Legendre function on the cut~\cite{gradstein}. Formula
(\ref{eq:GdS}) assumes $-1<Z<1$, and an appropriate analytic
continuation gives $G_{\mathrm{dS}}^{(m)}$ for other values of
$Z$~\cite{bousso}. By Fourier analysis in~$\varphi$, the Green's
function in the metric (\ref{CFTmetric}) thus reads
\begin{align}
G
(T,\hat{x},\varphi;T',\hat{x}',\varphi')
&=
\frac{1}{2\pi \lambda_+}
\int_{-\infty}^{\infty}
\rmd m \, \rme^{im(\varphi-\varphi')}
\, 
G_{\mathrm{dS}}^{(m)}
(T,\hat{x};T',\hat{x}')
\nonumber
\\
&=
\frac{\Gamma \bigl(\frac{d-3}{2} \bigr) }
{2{(2\pi)}^{(d-1)/2}}
\times 
\frac{1}{{\bigl[
\cosh(\lambda_{+}\Delta\varphi)
-Z(T,\hat{x};T',\hat{x}') \bigr]}^{(d-3)/{2}}}
\ \ , 
\nonumber
\\
\label{noident}
\end{align}
where $\Delta\varphi := \varphi - \varphi'$ and we have evaluated the
integral over $m$ using 7.217 in~\cite{gradstein}. The denominator in
(\ref{noident}) is positive for $Z < \cosh(\lambda_{+}\Delta\varphi)$,
and an appropriate analytic continuation gives $G
(T,\hat{x},\varphi;T',\hat{x}',\varphi')$ for other values of~$Z$. It
can be readily verified that (\ref{noident}) satisfies the wave
equation~(\ref{waveeq}). Note that conformally transforming
(\ref{noident}) to the Einstein static universe by
(\ref{CFTmetric})--(\ref{eq:ESU}) gives the Einstein static universe
Green's function in the vacuum that is positive frequency with respect
to~$\partial_\tau$.

The Green's function (\ref{noident}) is periodic in $T$ with period
$2\pi i$. This can be understood as a consequence of the periodicity
of the Euclidean vacuum Green's function on
dS${}_{d-2}$~\cite{gh-deS}. By the contour deformation argument
in~\cite{takagi}, the response function of an inertial monopole
particle detector at constant $\varphi$ satisfies the
Kubo-Martin-Schwinger (KMS) condition at temperature $1/(2\pi)$, and
our Euclidean vacuum is thus thermal in this sense. The contour
deformation argument does however not generalise to an inertial
particle detector that has velocity in the $\varphi$ direction, and we
have found no reason to expect the response of such a detector to be
thermal.

Finally, turn to the boundary of the \hdbtz{} hole, where
$\varphi$ has period~$2\pi$. The separation constant $m$ in 
the modes $\phi_{mnk}$ (\ref{eq:evac-modes-full}) then becomes an
integer. The complex analytic properties of the modes $\psi_{mnk}$ on the
Einstein static universe remain unchanged, and the Euclidean
vacuum is still that induced from the global vacuum on the conformal
boundary of~AdS${}_d$. The Green's function is obtained from
(\ref{noident}) by the method of images, with the result 
\begin{equation}
G(x;x')
=\frac{\Gamma \bigl(\frac{d-3}{2} \bigr) }
{2{(2\pi)}^{(d-1)/2}}
\times 
\sum_{k\in\BbbZ}
\frac{1}{{\bigl\{
\cosh \bigl[ \lambda_{+}(\Delta\varphi + 2\pi k) \bigr]
-Z(T,\hat{x};T',\hat{x}') \bigr\}}^{(d-3)/{2}}}
\ \ . 
\label{CFTG}
\end{equation}
The contour deformation argument of 
\cite{takagi} shows again that the response function of an inertial
monopole particle detector at constant $\varphi$ satisfies the KMS
condition at temperature~$1/(2\pi)$.

\subsection{Boundary Green's function from bulk geodesics}
\label{bulkBTZ}

In this subsection we show that the bulk geodesic approximation of 
\cite{Bala-Ross} reproduces the Euclidean vacuum Green's
function on the conformal boundary of the \hdbtz{} hole. 

Recall that in the geodesic method 
one first assumes for an appropriate bulk Green's function the estimate 
\begin{equation}
G_{\mathrm{bulk}}(y,y) \sim P \exp[-\mu L(y,y') ] 
\ \ , 
\label{eq:bulk-estimate} 
\end{equation}
where $L(y,y')$ is the length of a (spacelike) geodesic connecting the
bulk points $y$ and $y'$, $\mu$ is a constant and $P$ is a prefactor that is
slowly varying in some suitable sense. When
$y$ and
$y'$ tend to two spacelike-separated points
on the conformal boundary, $L(y,y')$ diverges, 
but a finite remainder may be obtained via multiplicative renormalisation of
(\ref{eq:bulk-estimate}) by a function of the
hypersurfaces through which 
$y$ and $y'$ approach the conformal boundary. The aim is to interpret this
remainder as a boundary Green's function of a conformal scalar field whose
conformal weight depends on~$\mu$. 

Consider now AdS${}_{d}$ in the 
parametrisation~(\ref{eq:exterior-para}), 
take $y = (T,\rho,\hat{x},\varphi)$, 
$y' = (T',\rho,\hat{x}',\varphi')$, and let $\rho\to\infty$. 
Writing 
$D:=-{(\Delta X^{0})}^{2}+{(\Delta X^{1})}^{2}+\cdots+{(\Delta
X^{d-1})}^{2}-{(\Delta X^{d})}^{2}$, a direct computation gives 
\begin{equation}
\frac{D(y,y')}{2\myell^2} 
= 
- 1 
+ 
\rho^2
\left[
\cosh \! \left( \lambda_+ \Delta\varphi \right)
- \left( 1 - \rho^{-2} \right)
\! 
Z 
( T , \hat{x} ; T' , \hat{x}' ) 
\right]
\ \ , 
\label{eq:Dyyprime}
\end{equation}
where 
$Z$ was defined in~(\ref{eq:Zdef}). 
From the symmetries of AdS${}_{d}$ 
it follows that the relation between $D$
and the geodesic distance $L$ is
\begin{equation}
\sinh^{2}
\! \left( \frac{L}{2\myell} \right)
=
\frac{D}{4\myell^2}
\ \ . 
\label{adsL}
\end{equation}
For $\mu>0$, 
(\ref{eq:Dyyprime})
and 
(\ref{adsL}) imply
\begin{equation}
\rme^{-\mu L} = 
{\left( 2 \rho^2 \right)}^{- \mu \myell} 
\times
\frac{1}{{\bigl[
\cosh(\lambda_{+}\Delta\varphi)
-Z(T,\hat{x};T',\hat{x}') \bigr]}^{\mu\myell}}
\times 
\left[1+\mathcal{O}
\left( \rho^{-2} \right)
\right]
\ \ . 
\label{bulkG}
\end{equation}
With the choice $\mu\myell = \tfrac12 (d-3)$, (\ref{bulkG}) thus gives
\begin{equation}
\frac{\Gamma \bigl(\frac{d-3}{2} \bigr) }
{4{\pi}^{(d-1)/2}}
\times
\rho^{2\mu\myell}
\, 
\rme^{-\mu L}
\xrightarrow[\rho\to\infty]{}
G (T,\hat{x},\varphi;T',\hat{x}',\varphi')
\ \ , 
\label{eq:geod-result}
\end{equation}
with $G (T,\hat{x},\varphi;T',\hat{x}',\varphi')$ given
by~(\ref{noident}). This shows that the geodesic approximation with
$\mu\myell = \tfrac12 (d-3)$ reproduces the Euclidean vacuum Green's
function on $\mathrm{dS}_{d-2} \times \BbbR$. Taking the periodic sum
in $\varphi$ reproduces the Euclidean vacuum Green's function on the
boundary of the
\hdbtz{} hole.

The geodesics involved in the result (\ref{eq:geod-result})
are real and spacelike 
for $\cosh ( \lambda_{+} \Delta\varphi )
-Z(T,\hat{x};T',\hat{x}') >0$, 
while for other values the geodesic 
approximation is understood 
in the sense of analytic continuation. 
As geodesics in de~Sitter space have 
$Z\ge-1$ \cite{allen-DSvacuum}, it follows from the 
$\rmd s^2_{\mathrm{dS}}$ term in (\ref{BTZ}) that 
the real spacelike geodesics 
for which 
$Z<-1$ have to pass inside the \hdbtz\ horizon.

\subsection{Boundary stress-energy}
\label{subsec:boundary-stress}

In AdS/CFT correspondence, one expects boundary stress-energy to be
related to the bulk mass, assuming a bulk mass can be independently
defined. We now examine this issue for the \hdbtz{} hole.

Consider the boundary. The symmetries of the metric (\ref{CFTmetric})
and the Green's function (\ref{CFTG}) imply that the Euclidean vacuum
stress-energy expectation value takes the form $\langle
T_{\mu\nu}\rangle = a {(g_{\mathrm{dS}})}_{\mu\nu} + b \bigl[
g_{\mu\nu} - {(g_{\mathrm{dS}})}_{\mu\nu} \bigr]$, where $a$ and $b$
are constants and $(g_{\mathrm{dS}})_{\mu\nu}$ is the
de~Sitter metric~(\ref{eq:dSglobalmetric}). 
To determine $a$
and~$b$, one starts from the classical expression for~$T_{\mu\nu}$,
which in our conventions reads~\cite{birrell}
\begin{align}
T_{\mu\nu}
&=
\frac{d-1}{2(d-2)}\phi,_{\mu}\phi,_{\nu}-\frac{1}{2(d-2)}
g_{\mu\nu}g^{\rho\sigma}\phi,_{\rho}\phi,_{\sigma}
\nonumber 
\\
&
- \frac{d-3}{2(d-2)}\phi_{;\mu\nu}
\phi 
+ \frac{{(d-3)}^{2}}{8(d-2)}
\bigl[ -g_{\mu\nu}+2{(g_{\mathrm{dS}})}_{\mu\nu} \bigr]
\phi^{2}
\ \ . 
\label{T-classical} 
\end{align} 
One point-splits~(\ref{T-classical}), reinterprets it in terms of
the Green's function, and finally takes the coincidence limit after
subtracting the divergent geometric
part~\cite{birrell}. 

The main computational effort in this point-splitting would be in the
divergent geometric part, which becomes increasingly complicated with
increasing dimension. As our principal interest is in the
$\lambda_+$-dependence of~$\langle T_{\mu\nu}\rangle$, we only
evaluate $\Delta T_{\mu\nu}$, the difference of $\langle
T_{\mu\nu}\rangle$ between the spacetimes with periodic and
nonperiodic~$\varphi$. $\Delta T_{\mu\nu}$ is the contribution from
the $k\ne0$ terms in (\ref{CFTG}) and requires no renormalisation. A
straightforward computation shows that the only nonvanishing
components are
\begin{subequations}
\label{stress123}  
\begin{align}
\Delta T^{T}_{T}
&=
\frac{\Gamma\bigl(\frac{d-1}{2}\bigr)}{4(d-2) (2\pi)^{(d-1)/2}}
\sum_{k\neq0}
\frac{(d-3) \cosh(2\pi k\lambda_{+}) + (d-1) }
{{\bigl[ \cosh(2\pi k\lambda_{+}) \bigr]}^{(d-1)/2}}
\ \ , 
\label{stress1}  
\\
\Delta T^{i}_{j}
&= \Delta T^{T}_{T}
\, \delta^{i}_{j}
\ \ , 
\label{stress2}
\\ 
\Delta T^{\varphi}_{\varphi}
&= - (d-2) \Delta T^{T}_{T}
\ \ , 
\label{stress3}
\end{align}
\end{subequations}
where the indices in (\ref{stress2}) are on $S^{d-3}$. Note that
$\Delta T_{\mu\nu}$ is traceless. 

Now, the energy density seen by an inertial observer at constant
$\varphi$ in the metric (\ref{CFTmetric}) is $E = E_0 -
\Delta T^{T}_{T}$, where $E_0$ is the $\lambda_+$-independent
contribution from the $k=0$ term in (\ref{CFTG}) after
renormalisation. In the limit of large~$\lambda_+$, the dominant terms in
(\ref{stress1}) are those with $k=\pm1$, and we find 
\begin{equation}
E 
\sim 
E_0
- 
\frac{ (d-3) \Gamma \bigl(\frac{d-1}{2}\bigr) }{4 (d-2) \pi^{(d-1)/2}}
\, \rme^{-(d-3)\pi\lambda_{+}}
\ \ . 
\label{approxstress1}  
\end{equation}
For a large
(and hence presumably classical) \hdbtz{} hole, $\lambda_+\gg1$, 
AdS/CFT correspondence thus suggests associating with the hole the CFT
mass
\begin{equation}
\Mcft
= 
A \left(  
1 - 
\rme^{-(d-3)\pi\lambda_{+}}
\right) 
\ \ , 
\label{eq:Mcft-def}
\end{equation}
where $A$ is some positive constant and 
we have normalised the zero of $\Mcft$
to $\lambda_+=0$. 

Consider then the bulk. As the neighbourhood of the infinity is not
asymptotically stationary, an ADM mass is not defined. For $d=5$, a
conserved charge proportional to $\lambda_+^2$ was identified by
embedding the bulk in Chern-Simons
supergravity~\cite{banados,gomberoff}. In comparison, the $d=3$ ADM
mass is proportional to $\lambda_+^2$~\cite{teitelboim,zanelli}. We
note that the relation (\ref{eq:Mcft-def}) between $\Mcft$ and
$\lambda_+$ resembles the relation between the ADM energy $H$ and
C-energy $c$ for cylindrical gravitational waves in four dimensions,
\begin{equation}
H = \frac{1}{4G} \left( 1 - \rme^{-4Gc} \right) 
\ \ , 
\end{equation}
where $G$ is Newton's constant~\cite{ash-vara}. It would be
interesting to understand whether this resemblance is more than a
coincidence.

An elementary analysis shows that the $k\ne0$ coincidence limit
geodesics are contained in the \hdbtz\ exterior and pass asymptotically
close to the horizon as $\lambda_+|k|\to\infty$. From the geodesic
approximation viewpoint, $\Delta T_{\mu\nu}$ does therefore not probe
the geometry behind the horizons but the result (\ref{approxstress1})
probes the near-horizon region of the exterior.

\subsection{$\BbbZ_2$ quotient}
\label{subsec:hdbtz-geon}

We next consider the quotient of the region 
$X^{d} > \abs{X^{d-1}}$ in
AdS${}_d$ under the group generated by the isometry $J := \exp(\pi
\lambda_+ \xi) \circ J_0$, where
\begin{equation}
J_0: 
\bigl( X^0, X^1, \cdots , X^{d-2}, X^{d-1}, X^{d} \bigr) 
\mapsto 
\bigl( X^0, - X^1, \cdots , - X^{d-2}, X^{d-1}, X^{d} \bigr) 
\ \ . 
\end{equation}
As $J^2 = \exp(2\pi \lambda_+ \xi)$, this spacetime
is a $\BbbZ_2$ quotient the \hdbtz{} hole. 
The construction resembles that of the 
$\RPthree$ and $\RPtwo$ geons as $\BbbZ_2$ quotients of respectively
Kruskal \cite{MW-geondata,giulini,FriedSchWi} and the nonrotating BTZ
hole~\cite{louko-marolf}, 
but while the $\RPthree$ and $\RPtwo$ geon quotients identify two
disconnected exterior regions, the exterior of the \hdbtz{} hole is already
connected, and the $\BbbZ_2$ identification in the exterior coordinates 
(\ref{eq:BTZ-fullset}) reads
\begin{equation}
(T,\rho,x^{i},\varphi)\sim(T,\rho,-x^{i},\varphi+\pi)
\ \ . 
\label{identification}
\end{equation}

The metric on the conformal boundary is 
given by (\ref{eq:dSglobalmetric}) and
(\ref{CFTmetric}) with the
identification 
\begin{equation}
(T,x^{i},\varphi) \sim (T,-x^{i},\varphi+\pi)
\ \ . 
\label{b-identification}
\end{equation}
Note that among the continuous de~Sitter isometries
in~(\ref{CFTmetric}), (\ref{b-identification}) preserves only the
rotations on~$S^{d-3}$. The boundary has thus a distinguished
spacelike foliation 
by the constant $T$ hypersurfaces. 
Similarly, the bulk exterior has a distinguished
spacelike 
foliation by the constant $T$ hypersurfaces in~(\ref{eq:BTZ-fullset}).

The Euclidean vacuum on the conformal boundary of the \hdbtz{} hole induces
a Euclidean vacuum on the conformal boundary of the $\BbbZ_2$ quotient. By
the method of images, the Green's function is the sum of 
(\ref{CFTG}) and the additional term 
\begin{align} 
&
\Delta_g G(x;x')
\nonumber
\\
& 
\ \ 
:=
\frac{\Gamma \bigl(\frac{d-3}{2} \bigr) }
{2{(2\pi)}^{(d-1)/2}}
\times 
\sum_{k\in\BbbZ}
\frac{1}{{\Bigl(
\cosh \bigl\{ \lambda_{+}[\Delta\varphi + (2 k + 1)\pi ] \bigr\}
-Z(T,\hat{x};T',-\hat{x}') \Bigr)}^{(d-3)/{2}}}
\ \ , 
\label{CFTG-geoncorr}
\end{align}
and it follows from subsection \ref{bulkBTZ} that the term
(\ref{CFTG-geoncorr}) is reproduced by the bulk geodesic approximation. 

The contribution from $\Delta_g G(x;x')$ 
(\ref{CFTG-geoncorr}) to the stress-energy tensor requires no
renormalisation. In the notation of~(\ref{stress123}), a direct computation
yields 
\begin{subequations}
\label{geonstress123}
\begin{align}
\Delta_g T^{T}_{T}
&=
\frac{(d-3)\Gamma
\bigl(\frac{d-1}{2}\bigr)}{4 (d-2) (2\pi)^{(d-1)/2}}
\sum_{k\in\mathbb{Z}}\frac{C_{k}+1}{{(C_{k}-Z)}^{{(d-1)/2}}}
\ \ , 
\label{geonstress1}
\\
\Delta_g T^{i}_{j}
&=
\delta^{i}_{j} 
\, 
\frac{\Gamma\bigl(\frac{d-1}{2}\bigr)}
{4 (d-2) (2\pi)^{(d-1)/2}}
\sum_{k\in\mathbb{Z}}\frac{(C_{k}+1)
\bigl[d(C_{k}-1)+2(Z-C_{k}) \bigr]}{{(C_{k}-Z)}^{{(d+1)/2}}}
\ \ ,   
\label{geonstress2}  
\\
\Delta_g T^{\varphi}_{\varphi}
&=
-\frac{(d-3)
\Gamma\bigl(\frac{d-1}{2}\bigr)}{4 (d-2) (2\pi)^{(d-1)/2}}
\sum_{k\in\mathbb{Z}}\frac{dC_{k}^{2}-1+(C_{k}+1)(Z-C_{k})}
{{(C_{k}-Z)}^{{(d+1)/2}}}
\ \ , 
\label{geonstress3}
\end{align}
\end{subequations}
where $C_{k}:=\cosh[(2k+1)\pi\lambda_{+}]$ and 
$Z = -\cosh(2T)$. 
Note that 
$\Delta_g T_{\mu\nu}$ is traceless. 

As the boundary is not globally de~Sitter invariant, there is no reason to
expect $\Delta_g T_{\mu\nu}$ to be locally de~Sitter invariant, and the
$T$-dependence in (\ref{geonstress123}) shows that it indeed is
not. We now show that the relative magnitude of 
$\Delta T_{\mu\nu}$ and $\Delta_g T_{\mu\nu}$ depends on both $\lambda_+$
and~$T$. 

Consider first the limit of large~$\lambda_+$ with fixed~$T$. The
leading terms in
$\Delta_g T_{\mu\nu}$ (\ref{geonstress123}) are those with
$k=0$ and $-1$, proportional to $\exp
\bigl[-\tfrac12 (d-3)\pi\lambda_{+} \bigr]$, and the
$T$-dependence only appears in a subleading order. In comparison, the
leading terms in $\Delta T_{\mu\nu}$ (\ref{stress123}) are proportional to
$\exp
\bigl[-(d-3)\pi\lambda_{+} \bigr]$. Hence $\Delta_g T_{\mu\nu}$ dominates
$\Delta T_{\mu\nu}$. 

Consider then the limit of large $\abs{T}$ with
fixed~$\lambda_+$. In~(\ref{geonstress123}), we first arrange the sums
in the form $\sum_{k=0}^{\infty} {(C_k-Z)}^{-p}$ with $p>0$ and use the
results in appendix \ref{app:b-asymptotics} to 
identify the leading sums and 
replace $C_k$ by $\tfrac12
{\rme}^{\pi\lambda_{+}} {\rme}^{2\pi k \lambda_{+}}$. 
We then
rearrange the leading sums into sums of the form 
\begin{equation}
\sum_{k=0}^{\infty}
\frac{
\tfrac12 {(-Z)}^{-1} 
{\rme}^{\pi\lambda_+} {\rme}^{2\pi k \lambda_+}
}
{
{\bigl[ 
\tfrac12 {(-Z)}^{-1} 
{\rme}^{\pi\lambda_+} {\rme}^{2\pi k\lambda_{+}}
+1
\bigr]}^{p}}
\ \ , \ \ 
p>1
\ \ . 
\label{eq:semi-sumZ}
\end{equation} 
In the sums~(\ref{eq:semi-sumZ}), we next include also negative
integer values of~$k$: This introduces an error of order
$\mathcal{O}\bigl(Z^{-1}\bigr)$, but as the new sums are periodic
in~$\ln(-Z)$, the error is sub-leading.  Finally, replacing $Z$ by
$-\tfrac12 \rme^{2 \abs{T}}$, we find the asymptotic large $\abs{T}$
behaviour
\begin{subequations}
\label{geonstress-largeT}
\begin{align}
\Delta_g T^{T}_{T}
&\sim
\frac{(d-3)\Gamma \bigl( \frac{d-1}{2}
\bigr)}{4 (d-2) \pi^{(d-1)/2}}
\, 
{\rme^{-(d-3)\abs{T}}}
\, 
f_{(d-1)/2} 
\ \ , 
\\ 
\Delta_g T^{i}_{j}
&\sim 
\delta^{i}_{j}
\, 
\frac{\Gamma \bigl( \frac{d-1}{2}
\bigr)}{4 (d-2) \pi^{(d-1)/2}}
\, 
{\rme^{-(d-3)\abs{T}}}
\, 
\bigl[ 
(d-3)f_{(d-1)/2} 
- (d-1) f_{(d+1)/2} 
\bigr]
\ \ , 
\\ 
\Delta_g
T^{\varphi}_{\varphi}
&\sim 
-
\frac{(d-3) \Gamma \bigl( \frac{d-1}{2}
\bigr)}{4 (d-2) \pi^{(d-1)/2}}
\, 
{\rme^{-(d-3)\abs{T}}}
\, 
\bigl[ 
(d-2)f_{(d-1)/2} 
- (d-1) f_{(d+1)/2} 
\bigr]
\ \ , 
\end{align}
\end{subequations}
where 
\begin{equation}
f_{p} :=\sum_{k\in\mathbb{Z}}
\frac{{\rme}^{(2k+1)\pi\lambda_{+}-2\abs{T}}}
{{\bigl[{\rme}^{(2k+1)\pi\lambda_{+}-2\abs{T}}+1\bigr]}^{p}}
\ \ , \ \ 
p>1
\ \ . 
\end{equation} 
Note that $f_{p}$ is periodic in~$\abs{T}$, with period~$\pi\lambda_+$, 
and hence bounded in~$\abs{T}$.\footnote{Viewed as a function of
complexified~$\abs{T}$, $f_{p}$ has also period $\pi i$ and is hence an
elliptic function. For $p=2$, it can be expressed in terms of the
Weierstrass elliptic function (\cite{lang-elliptic}, pp.\ 45--47).}
Equations (\ref{geonstress-largeT}) thus show that
$\Delta_g T_{\mu\nu}$ decays exponentially 
as $\abs{T}\to\infty$. 

For given $\lambda_+\gg1$, these results imply that $\Delta_g
T_{\mu\nu}$ dominates $\Delta T_{\mu\nu}$ for some finite interval in
$T$ but decays exponentially as $\abs{T}\to\infty$. This decay could be
expected from similar results in geon-type versions of Rindler space
\cite{louko-marolf-rp3,langlois-rp3} and in the $\RPthree$ version of
de~Sitter space~\cite{louko-schleich}. 
Note that the dominance of $\Delta_g T_{\mu\nu}$ does not contradict the
proposal (\ref{eq:Mcft-def}) for the mass of the \hdbtz{} hole, as the
infinity neighbourhoods in the
\hdbtz{} hole and the $\BbbZ_2$ quotient are not globally isometric. 
Whether $\Delta_g T_{TT} + \Delta T_{TT}$ might provide a reasonable mass
for the $\BbbZ_2$ quotient is not clear, but the 
$T$-dependence of such a mass would not contradict any bulk symmetries. 

Finally, note that all the new $T\ne0$ coincidence limit geodesics have
$Z<-1$ and hence pass inside the horizon. From the geodesic
approximation viewpoint this means that $\Delta_g T_{\mu\nu}$ arises
entirely from geodesics that probe the geometry behind the horizon.

\subsection{Adding rotation to the \hdbtz{} hole}
\label{subsec:rotHDBTZ}

The \hdbtz{} hole is a direct generalisation of the nonrotating BTZ
hole to $d\ge4$. Generalising the rotating BTZ hole in
a similar way produces a spacetime that is not a black
hole~\cite{holst}, but part of the conformal boundary of this
spacetime is still obtained from the metric (\ref{CFTmetric}) by an
appropriate identification, and the methods of subsections
\ref{subsec:euclidean} and
\ref{bulkBTZ} 
define on this part of the boundary a conformal scalar
field state. We now discuss briefly the stress-energy of this state in view
of AdS/CFT correspondence.

In the bulk, the isometry group is now generated by $\exp\bigl( 2\pi
\lambda_+\xi + \lambda_-\eta \bigr)$, where $\xi$ is as
in~(\ref{eq:xi-def}),
\begin{equation}
\eta := 
X^{0}
\frac{\partial}{\partial{X^{d-2}}} 
+ X^{d-2} 
\frac{\partial}{\partial{X^{0}}}
\ \ , 
\label{eq:eta-def}
\end{equation}
and the parameters $\lambda_+$ and $\lambda_-$ satisfy $0< \lambda_- <
2\pi\lambda_+$. The region of interest on the conformal boundary without
identifications is (\ref{CFTmetric}) with $-\infty < \varphi
< \infty$, and the group acting on it is generated by $\exp\bigl( 2\pi
\partial_\varphi + \lambda_-\tilde\eta \bigr)$, where $\tilde\eta$ is
induced by~$\eta$: $\tilde\eta$ is a boost-like Killing vector on 
$\rmd s^2_{\mathrm{dS}}$ in~(\ref{CFTmetric}), 
normalised so that 
${\tilde\eta}^\mu {\tilde\eta}_\mu \ge -1$. Note that $2\pi
\partial_\varphi + \lambda_-\tilde\eta$ is spacelike. 
In a static coordinate patch
$(\tilde{t},\tilde{r},\tilde{x}^{i},\varphi)$ adapted to~$\tilde\eta$,
the metric (\ref{CFTmetric}) reads 
\begin{equation}
\rmd s^{2}_{\mathrm{CFT}}
=
-(1-\tilde{r}^{2})
\rmd \tilde{t}^{2}+ {(1-\tilde{r}^{2})}^{-1} \rmd \tilde{r}^{2}
+\tilde{r}^{2}\rmd\tilde{\Omega}_{d-4}^2
+\lambda_{+}^{2}\rmd\varphi^{2} 
\ \ , 
\label{eq:static-coords}
\end{equation}
where $0\le \tilde{r} <1$, 
$\tilde\eta = \partial_{\tilde{t}}$, and the identification
is 
\begin{equation}
(\tilde{t},\tilde{r},\tilde{x}^{i},\varphi)
\sim
(\tilde{t}+\lambda_{-},\tilde{r},\tilde{x}^{i},\varphi+2\pi)
\ \ , 
\end{equation}
where $\tilde{x}$ is the $(d-3)$-dimensional unit vector
coordinatising~$S^{d-4}$. 

The Green's function is constructed from (\ref{noident}) by the method
of images, and the contribution to the boundary stress-energy tensor
from the identifications can be computed as in
subsection~\ref{subsec:boundary-stress}. 
Working in the static
coordinates~(\ref{eq:static-coords}), the differentiations can be
performed with the help of the formula
\begin{equation}
Z(x;x')=\sqrt{(1-\tilde{r}^{2})(1-
{\vphantom{(}\tilde{r}'\vphantom{)}}^{2}
)}
\cosh(\Delta\tilde{t})+\tilde{r}\tilde{r}'
\sum_{i=1}^{d-3}\tilde{x}^{i}
{\vphantom{(}\tilde{x}'\vphantom{)}}^{i}
\ \ . 
\end{equation}
In the limit of large
$\lambda_+$ with fixed $\lambda_-/\lambda_+$, the leading asymptotic
behaviour is 
\begin{subequations}
\label{eq:spin-boundary-T} 
\begin{align}
\Delta
T^{\tilde{t}}_{\tilde{t}}
&\sim
F
\left[(d-3) + (d-1 -2\tilde{r}^{2}) q 
+(d-2)(d-3) (1-\tilde{r}^{2}) q^2
\right] 
\ \ , 
\\
\Delta
T^{\tilde{r}}_{\tilde{r}}
&\sim
F (d-3)
\left[1 + \tilde{r}^{2} q 
- (1-\tilde{r}^{2})q^2 
\right] 
\ \ , 
\\
\Delta
T^{\tilde i}_{\tilde j}
&\sim
F 
\left[(d-3)
-2\tilde{r}^{2} q 
-(d-3)(1-\tilde{r}^{2}) q^2 
\right] 
\ \ , 
\\
\Delta
T^{\varphi}_{\varphi}
&\sim
- F 
\left[ (d-2)(d-3)
+\left((d-3)(1 - \tilde{r}^{2}) + 2\right) q 
+ (d-3)(1-\tilde{r}^{2}) q^2 
\right] 
\ \ , 
\\
\Delta
T^{\tilde{t}}_{\varphi}
&\sim
- F 
\lambda_{+} d (d-2) q 
\ \ , 
\end{align}
\end{subequations}
where 
\begin{subequations}
\begin{align}
F 
&:= 
\frac{\Gamma\bigl(\frac{d-1}{2}\bigr) \rme^{-(d-3)\pi\lambda_{+}} }
{ 4 (d-2) \pi^{(d-1)/{2}}
{\bigl[ 1 - (1-\tilde{r}^{2}) q \bigr]}^{(d+1)/{2}}
}
\ \ , 
\\
q 
&:= 
\rme^{\lambda_- - 2\pi\lambda_+}
\ \ . 
\end{align}
\end{subequations}
Note that $\Delta T^{\tilde{t}}_{\varphi}$ is nonvanishing, but at large
$\lambda_+$ with fixed $\lambda_-/\lambda_+$ it is exponentially suppressed
compared with the diagonal components.

Now, AdS/CFT correspondence suggests seeking in the boundary
stress-energy (\ref{eq:spin-boundary-T}) evidence of rotation
in the bulk spacetime, and possibly even a definition of the angular
momentum of the bulk spacetime. While the nonvanishing value of $\Delta
T^{\tilde{t}}_{\varphi}$ can qualitatively be seen as such evidence,
it is difficult to identify a more quantitative correspondence. The
nondiagonal component has a nontrivial dependence on $\tilde{r}$ even
when expressed in the normalised basis, and we have verified that the
same holds also in a Lorentz-orthonormal basis in which the spacelike
unit vector in $\mathrm{span}\{\partial_{\tilde{t}},
\partial_\varphi\}$ is proportional to $2\pi
\partial_\varphi + \lambda_-\tilde\eta$.

\section{Black holes with flat boundaries}
\label{sec:schw-flat}

In this section we investigate spacetimes obtained from the
Schwarzschild-AdS family in $d\ge4$ dimensions by replacing the round
$S^{d-2}$ by a flat space
\cite{GibbWilt,lemos1,lemos2,lemos-zanchin,birmingham-topol}. The metric in
the curvature coordinates $(x^{\mu})=(t,r,\vec{x})$ reads
\begin{subequations}
\label{eq:gwmetric}
\begin{align}
\rmd s^{2} 
& = 
-f \rmd t^{2} + \frac{\rmd r^{2}}{f}  
+ \frac{r^{2}}{\myell^2} 
\, \rmd\vec{x}^{2}
\ \ ,
\label{eq:gw-metric}
\\
f(r) 
& := 
\frac{r^{2}}{\myell^{2}} - \frac{2M}{r^{d-3}}
\ \ , 
\end{align}
\end{subequations}
where $\vec{x}^{2} := \sum_{i=1}^{d-2}
(x^{i})^{2}$. The positive parameter $\myell$ is related to the
cosmological constant by $\Lambda = - \tfrac12 (d-1)(d-2)
\myell^{-2}$, and the positive parameter $M$ is proportional to the
ADM mass per unit coordinate volume
in~$\rmd\vec{x}^{2}$~\cite{chrusciel,nagy}. The value of $r$ at the
horizon, where $f$ vanishes, is $r_{h} :=
\sqrt[d-1]{2M\myell^{2}}$. The surface gravity of the horizon,
normalised to the Killing vector~$\partial_t$, is $\kappa =
\tfrac12 (d-1)r_h \myell^{-2}$, and the inverse Hawking temperature is
$\beta_H := 2\pi/\kappa = \tfrac{4\pi}{(d-1)} \myell^2 r_h^{-1}$. 

We take initially $\vec{x} \in \BbbR^{d-2}$, so that the horizon has
(spatial) topology~$\BbbR^{d-2}$. The parameter $M$ has then
coordinate-invariant meaning only in being positive, as its value can be
rescaled by rescaling $r$, $t$ and~$\vec{x}$. Other horizon topologies
will be discussed in subsections \ref{subsec:periodic}
and~\ref{subsec:sads-geon}. 

The metric on hypersurfaces of constant $r$ satisfies 
\begin{equation}
\left. 
\rmd s^{2}
\right|_{r=\mathrm{const}}
\sim
\frac{r^{2}}{\myell^2} 
\bigl(
-\rmd
t^{2} + \rmd \vec{x}^{2}
\bigr)
\ \ , 
\ \ 
r\to\infty
\ \ , 
\end{equation}
and a convenient 
representative of the conformal equivalence class of metrics at
the infinity is the $(d-1)$-dimensional Minkowski metric, 
\begin{equation}
\rmd s^{2}_{\mathrm{CFT}}
=
-\rmd
t^{2} + \rmd \vec{x}^{2}
\ \ . 
\label{gw-CFTmetric}
\end{equation}
By AdS/CFT correspondence, one expects the bulk to induce on the boundary
(\ref{gw-CFTmetric}) a thermal state 
with inverse temperature $\beta_H$ and
energy expectation value proportional to~$M$. 
The 
periodicity of $t$ on the
Euclidean-signature section of the bulk implies that the boundary two-point
function evaluated in the geodesic approximation is 
periodic in $t$ with period~$i\beta_H$, but the
boundary stress-energy requires an explicit computation.
We now embark on this computation.

\subsection{Bulk geodesics}
\label{flatbulk}

It is convenient to analyse the bulk geodesics first on the
Euclidean-signature section and at the end continue to Lorentz-signature. 
We write
$t=-i\tau$ and regard $\tau$ in this subsection as real. 

The geodesic equations read 
\begin{subequations} 
\label{eq:txL-diff}
\begin{align}
\frac{\rmd \tau}{\rmd \sigma} 
& = 
\frac{C}{f(r)}
\ \ , 
\label{t}
\\
\frac{\rmd \vec{x}}{\rmd \sigma} 
& = 
\frac{ \myell \vec{C}}{r^{2}}
\ \ ,
\label{x}
\\
\left( \frac{\rmd r}{\rmd \sigma} \right)^2 
& = 
\frac{P(r)}{\myell^2 r^{d-1}} 
\ \ , 
\end{align}
\end{subequations} 
where $\sigma$ is the proper distance, 
$C$ and $\vec{C}$ are constants, and 
\begin{equation}
P(r)
:=
r^{d+1}
- \bigl( C^{2}\myell^{2}
+\vec{C}^{2} \bigr) r^{d-1}
-2M\myell^{2}r^{2}+2M\myell^{2}\vec{C}^{2} 
\ \ . 
\label{poly}
\end{equation}
We are interested in geodesics that begin and end at $r=\infty$ and have a
turning point at $r=\rmin > r_h$, which is the largest zero of~$P(r)$.
We truncate these geodesics at $r = \rmax$, where the cutoff $\rmax$ will
shortly be taken to infinity. Denoting by 
$\Delta\tau$ and $\Delta\vec{x}$ the differences in respectively 
$\tau$
and $\vec{x}$ between the endpoints of the truncated geodesic and by $L$
the length of the truncated geodesic, (\ref{eq:txL-diff}) gives 
\begin{subequations}
\begin{align}
\Delta \tau
& = 
2C \myell^{3} \int^{\rmax }_{\rmin } 
\frac{r^{\frac{3d-7}{2}} \rmd r}{\bigl( r^{d-1}-2M
\myell^{2} \bigr)\sqrt{P(r)}}
\ \ , 
\label{dt}
\\
\Delta \vec{x} 
& = 
2\vec{C}\myell^2 \int^{\rmax }_{\rmin } 
\frac{r^{\frac{d-5}{2}} \rmd r}{\sqrt{P(r)}}
\ \ , 
\label{dx}
\\
L 
& = 
2 \myell \int^{\rmax }_{\rmin } \frac{r^{\frac{d-1}{2}} \rmd
r}{\sqrt{P(r)}}
\ \ . 
\label{L}
\end{align}
\end{subequations}

Consider now the limit $\rmax\to\infty$ with fixed 
$C$ and~$\vec{C}$. $\Delta \tau$~and $\Delta \vec{x}$ tend to the finite
values 
\begin{subequations}
\label{eq:Ltx-ren}
\begin{align}
\Delta \tau
& = 
2C \myell^{3} \int^{\infty}_{\rmin } 
\frac{r^{\frac{3d-7}{2}} \rmd r}{\bigl(r^{d-1}-2M
\myell^{2}\bigr)\sqrt{P(r)}}
\ \ , 
\label{dt-ren}
\\
\Delta \vec{x} 
& = 
2\vec{C}\myell^2 \int^{\infty }_{\rmin } 
\frac{r^{\frac{d-5}{2}} \rmd r}{\sqrt{P(r)}}
\ \ . 
\label{dx-ren}
\end{align}
$L$ diverges, but subtracting from (\ref{L}) the
integral of $2 \myell {(r^2 - \rmin^2)}^{-1/2}$ and performing the
elementary integration shows that $\Lren := \lim_{\rmax\to\infty}
\bigl[ L - 2\myell \ln(\rmax/\myell) \bigr]$ 
is finite and given by 
\begin{equation}
\Lren
= 
{2\myell} \ln \! \left(\frac{2\myell}{\rmin} \right) 
+ 
{2\myell} 
\int^{\infty}_{\rmin
} 
\left( \frac{r^{\frac{d-1}{2}}}{\sqrt{P(r)}}
- 
\frac{1}{\sqrt{r^2 - \rmin^2}}
\right) 
\rmd r
\ \ .
\label{eq:Lren-def}
\end{equation}
\end{subequations}
Recalling that $\rmin$ is determined by $C$ and~$\vec{C}^2$, we see that the
system (\ref{eq:Ltx-ren}) determines $\Lren$ at least locally 
as a function of $\Delta\tau$
and~$\Delta\vec{x}$. We adopt this $\Lren$ as the renormalised
geodesic length to be used in the boundary Green's
function. 

To evaluate the boundary Green's function to quadratic order in
$\Delta\tau$ and~$\Delta\vec{x}$, which is the order that determines
the boundary stress-energy, it turns out sufficient to find the
expansion of $\Lren$ to
the next-to-leading order in $\Delta\tau$ and~$\Delta\vec{x}$. 
We show in appendix \ref{app:L-expansion} that this expansion is 
\begin{equation}
\Lren \sim 
2\myell \ln (D/\myell) +
\frac{ \sqrt{\pi} \, \Gamma \bigl( \frac{d-1}{2} \bigr)}{\Gamma
\bigl(\frac{d+2}{2}
\bigr)}
\times
\frac{MD^{d-3}}{2^{d}\myell^{2d-5}} 
\left[(\Delta\vec{x})^{2}-(d-2)(\Delta \tau)^{2}\right]
\ \ , 
\label{length}
\end{equation} 
where $D := \sqrt{(\Delta \tau)^{2} + (\Delta \vec{x})^{2}}$.

\subsection{Boundary CFT}
\label{subsec:flat-boundaryCFT}

We define on the Euclidean-signature section of the conformal boundary
(\ref{gw-CFTmetric}) the Green's function 
\begin{equation}
G_{\mathrm{CFT}} 
:= 
\frac{\Gamma \bigl(\frac{d-3}{2} \bigr) }
{4{\pi}^{(d-1)/2} \myell^{d-3}}
\, 
\rme^{-(d-3) \Lren/(2\myell)} 
\ \ . 
\label{eq:G-eucl-def}
\end{equation}
From~(\ref{length}), the expansion of $G_{\mathrm{CFT}}$ to
quadratic order in $\Delta\tau$ and~$\Delta\vec{x}$ reads 
\begin{align}
G_{\mathrm{CFT}}^{(2)}
&=
\frac{\Gamma \bigl(\frac{d-3}{2} \bigr) }
{4{\pi}^{(d-1)/2} }
\times 
\nonumber
\\
& 
\ \ 
\times 
\left\{
\frac{1}{D^{d-3}} 
+
\left[ \frac{2\pi}{(d-1)\beta_H} \right]^{d-1}
\frac{ (d-3) \sqrt{\pi} \, \Gamma \bigl( \frac{d-1}{2} \bigr)}{8 \Gamma
\bigl(\frac{d+2}{2}
\bigr)}
\left[ (d-2)(\Delta \tau)^{2} - (\Delta\vec{x})^{2} \right]
\right\}
\ \ , 
\label{eq:G-eucl-exp}
\end{align}
where the superscript ${}^{(2)}$ indicates that only terms up to quadratic
order in the coordinate separation 
have been kept. We have written $M$ in terms of the inverse Hawking
temperature~$\beta_H$. 
The coefficient of
$\Lren$ and the overall factor in (\ref{eq:G-eucl-def}) have been chosen so
that $G_{\mathrm{CFT}}$ has the short distance divergence of
a free
scalar field~\cite{birrell}. 

As $G_{\mathrm{CFT}}^{(2)}$ satisfies the Klein-Gordon equation, 
we can consistently view it as a free conformal scalar field
Green's function and compute the stress-energy from the
quadratic term. The
classical expression for the stress-energy is now given by the first three
terms in~(\ref{T-classical}), with $g_{\mu\nu}$ replaced by the Minkowski
metric~(\ref{gw-CFTmetric}). 
Continuing $G_{\mathrm{CFT}}^{(2)}$ to 
Lorentz-signature by $\tau=it$, the point-split calculation is
straightforward and yields 
\begin{equation}
{\langle T_{\mu\nu} \rangle}_{\!\mathrm{CFT}}
= 
\frac{ \left[ \Gamma \bigl( \frac{d-1}{2} \bigr)
\right]^2}
{8 {\pi}^{(d-2)/2}
\Gamma
\bigl(\frac{d+2}{2}
\bigr)}
\left[ \frac{2\pi}{(d-1)\beta_H} \right]^{d-1}
\times 
{\mathrm{diag}} ( d-2,1,\ldots,1 )
\ \ .
\label{eq:Tren-flathole}
\end{equation}
Note that ${\langle T_{\mu\nu} \rangle}_{\!\mathrm{CFT}}$ is traceless.
Expressing $\beta_H$ in terms of $M$ shows that (\ref{eq:Tren-flathole}) is
proportional to~$M$. The boundary energy density is therefore proportional
to the black hole ADM mass. This is the result one would
have expected. 

It is of interest to compare ${\langle T_{\mu\nu} \rangle}_{\!\mathrm{CFT}}$
(\ref{eq:Tren-flathole}) to the stress-energy ${\langle T_{\mu\nu}
\rangle}_{\mathrm{free}}$ of a free conformal scalar field on Minkowski
space in an ordinary thermal state with inverse temperature~$\beta_H$. From
appendix~\ref{perMink}, first term in
equation~(\ref{eq:perMink-T-leading}), we see that the expressions agree up
to a $d$-dependent multiplicative constant. 
We have verified by a combination of 
numerical methods at small $d$ and an asymptotic expansion at large $d$
that 
${\langle T_{00} \rangle}_{\!\mathrm{CFT}} / {\langle T_{00}
\rangle}_{\mathrm{free}}$ is less
than 1 for all integers $d >3$ (although it equals 1 at
$d\approx 3.1475$ and approaches 1 as $d\to3$) 
and decreases rapidly as $d$ increases.

\subsection{A periodic boundary dimension}
\label{subsec:periodic}

In this subsection we make one of the spatial dimensions in the bulk
metric (\ref{eq:gwmetric}) periodic by $\bigl(
t,r,x^1, x^2,\ldots,x^{d-2} \bigr) \sim \bigl( t,r,x^1 +a,
x^2,\ldots,x^{d-2} \bigr)$, where $a$ is a positive parameter. The
conformal boundary (\ref{gw-CFTmetric}) becomes then Minkowski space
with one periodic spatial dimension. We assume the extension past
the horizon to be of the usual Kruskal-type, reviewed in
appendix~\ref{app:kruskal}. This guarantees that 
the Lorentz-signature continuation of $G_{\mathrm{CFT}}$
(\ref{eq:G-eucl-def}) will receive no contribution from 
geodesics that cross the
horizon. 

The stress-energy of a free conformal scalar field in an
ordinary thermal state on Minkowski space with one periodic dimension,
computed in appendix~\ref{perMink}, suggests that in the limit of
large $a$ with fixed $M$ ${\langle T_{\mu\nu}
\rangle}_{\!\mathrm{CFT}}$ (\ref{eq:Tren-flathole}) should receive a
correction proportional to
\begin{equation}
\frac{1}{\beta_H a^{d-2}}
\times {\mathrm{diag}} ( 0, 3-d ,1,\ldots,1 )
\ \ .
\label{eq:ex-percorr}
\end{equation}
We now show that this suggestion is not realised.

A first observation is that any new contribution to ${\langle
T_{\mu\nu}
\rangle}_{\!\mathrm{CFT}}$ must be suppressed in $a$ exponentially, 
rather than as a power-law. In~(\ref{eq:Ltx-ren}), the
coincidence limit of the new geodesics occurs when 
$\Delta\tau = \Delta x^2 = \cdots = \Delta x^{d-2} = 0$ but $\Delta
x^1 = n a$, $n \in \BbbZ\setminus\{0\}$. This implies $C
=0$, $\vec{C}^2 = \rmin^2$ and $P(r) = (r^2 - \rmin^2)
\bigl( r^{d-1} - r_h^{d-1} \bigr)$. 
The limit of large $\abs{n} a$ is the limit
$\rmin \to r_h$, in which (\ref{dx-ren}) and (\ref{eq:Lren-def}) give 
\begin{subequations}
\label{eq:per-smallrmin}
\begin{align}
\frac{\abs{n}a}{\myell}  
& \sim
- \sqrt{\frac{2}{d-1}}
\, 
\frac{\myell}{r_h} 
\ln \! \left( \frac{\rmin}{r_h} -1 \right) 
\ \ , 
\\
\frac{\Lren}{\myell}
&\sim 
- \sqrt{\frac{2}{d-1}}
\ln \! \left( \frac{\rmin}{r_h} -1 \right) 
\ \ , 
\end{align}
\end{subequations}
and hence $\Lren \sim (r_h/\myell) \abs{n}a$. The dominant 
correction in $G_{\mathrm{CFT}}^{(2)}$  
thus comes from geodesics close to the 
coincidence 
geodesics with $n=\pm1$, and this contribution involves the
suppression factor $\rme^{-(d-3) \Lren/(2\myell)} \sim 
\exp \bigl[ - \tfrac12 (d-3) r_h a \myell^{-2}
\bigr]$. 

Computing the correction in $G_{\mathrm{CFT}}^{(2)}$ requires an
expansion around the coincidence geodesics with $n=\pm1$. What
makes the calculation laborious is that the leading contribution from
each geodesic is linear in the coordinate separation, and the linear
terms only cancel on adding the contributions from $n=1$ and
$n=-1$. We have only performed this calculation for $d=5$, in which
case the integrals in (\ref{eq:Ltx-ren}) simplify by taking $r^2$ as
the new variable.\footnote{The integrals can be evaluated
in terms of elliptic integrals. We have however not been able to
exploit this observation in the calculations.} Omitting here the
steps, we find that the correction in 
$G_{\mathrm{CFT}}^{(2)}$ reads
\begin{equation}
\frac{\Gamma \bigl(\frac{d-3}{2} \bigr) }
{4{\pi}^{(d-1)/2} }
\times 
\frac{{\bigl(1+\sqrt{2}\bigr)}^2}{4} 
\exp \bigl( - r_h a \myell^{-2} \bigr)
\frac{r_h^4}{\myell^8} 
{\bigl(\delta x^1\bigr)}^2
\ \ , 
\label{eq:per-corrgreen}
\end{equation}
where $\delta x^1$ denotes the separation in periodically
identified~$x^1$. The
terms involving $\Delta\tau$ and $\Delta x^i$ for $2\le i \le d-2$
are subleading in~$a$.

As (\ref{eq:per-corrgreen}) does not
satisfy the Klein-Gordon equation, we cannot interpret it as a
correction in a free conformal scalar field Green's function. We conclude
that our AdS/CFT correspondence model cannot be used to
compute the periodic correction to the boundary stress-energy at least for
$d=5$.

\subsection{Geon} 
\label{subsec:sads-geon}

Similarly to the $\RPthree$ extension of Schwarzschild
\cite{MW-geondata,giulini,FriedSchWi,louko-marolf-rp3} 
and the $\RPtwo$ extension of
nonrotating~BTZ~\cite{louko-marolf}, it is possible to continue the
exterior metric (\ref{eq:gwmetric}) with periodic $x^1$ 
into inextendible spacetimes that do not contain a second
exterior~\cite{LouMannMar}. The interest of these unconventional extensions
for our AdS/CFT correspondence model is that the Lorentz-signature
continuation of
$G_{\mathrm{CFT}}$ (\ref{eq:G-eucl-def}) may then receive additional
contributions from geodesics that cross the horizon. In the $\RPtwo$ case
such geodesics are required to recover the appropriate Green's
function on the boundary~\cite{lou-ma-ross}, and we observed a similar
phenomenon in the \hdbtz{} case in subsection~\ref{subsec:hdbtz-geon}. We
now address this question with the metric~(\ref{eq:gwmetric}). 

We consider the spacetime obtained from the Kruskal-type extension of
appendix \ref{app:kruskal} by the identification $\bigl( U, V, x^1,
x^2,\ldots,x^{d-2} \bigr) \sim \bigl( V,U,x^1 +
\tfrac12 a, x^2,\ldots,x^{d-2} \bigr)$. The global structure differs
from that of the $\RPtwo$ geon~\cite{louko-marolf} by the $d-3$
dimensions that are inert in the identification, and also by the fact
that the singularities in the conformal diagram cave inward compared
with the infinities (cf.~\cite{Fid-etal}). The exterior is as in
subsection~\ref{subsec:periodic}, but there are now horizon-crossing
geodesics that begin and end at the conformal boundary. 

Let $t=0$ be the distinguished constant $t$ hypersurface in the
exterior~\cite{louko-marolf,louko-marolf-rp3}. By the results in
subsection~\ref{subsec:hdbtz-geon}, one expects the contribution from
the horizon-crossing geodesics to be exponentially suppressed as
$\abs{t}\to\infty$ with fixed~$a$. 
However, at sufficiently small~$\abs{t}$,
the horizon-crossing geodesics dominate those considered in
subsection~\ref{subsec:periodic}. To see this, consider the
horizon-crossing coincidence limit geodesics at $t=0$. These geodesics
belong to the Euclidean-signature section at $\tau=0$ and are obtained
from (\ref{eq:Ltx-ren}) with 
$C=0$, 
$\rmin = r_h$, 
$\vec{C}^2 < r_h^2$, 
$P(r) = \bigl( r^2 - \vec{C}^2 \bigr)
\bigl( r^{d-1} - r_h^{d-1} \bigr)$, 
$\Delta\tau = \Delta x^2 = \cdots = \Delta x^{d-2} = 0$ and 
$\Delta x^1 = \bigl( \tfrac12 + n \bigr) a$, $n \in \BbbZ$. 
In parallel with~(\ref{eq:per-smallrmin}), we now find 
$\Lren \sim (r_h/\myell) \bigl\lvert \tfrac12 + n \bigr\rvert a$. 
The dominant 
correction in $G_{\mathrm{CFT}}^{(2)}$ 
thus comes from geodesics close to the 
coincidence 
geodesics with $n=0$ and $n=-1$ and involves the suppression factor 
$\rme^{-(d-3) \Lren/(2\myell)} \sim 
\exp \bigl[ - \tfrac14 (d-3) r_h a \myell^{-2}
\bigr]$, which goes to zero less rapidly than the factor
$\exp \bigl[ - \tfrac12 (d-3) r_h a \myell^{-2} \bigr]$
found in subsection~\ref{subsec:periodic}. 

We have computed the leading correction in $G_{\mathrm{CFT}}^{(2)}$
only for $d=5$, and then only at $t=0$. Written on the Lorentz-signature
section in the notation of~(\ref{eq:per-corrgreen}), we find that this
correction is 
\begin{equation}
\frac{\Gamma \bigl(\frac{d-3}{2} \bigr) }
{4{\pi}^{(d-1)/2} }
\times 
\frac{{\bigl(1+\sqrt{2}\bigr)}^2}{4} 
\exp \bigl( - \tfrac12 r_h a \myell^{-2} \bigr)
\frac{r_h^4}{\myell^8} 
\left[ 
{\bigl(\delta x^1\bigr)}^2 
+ 
3\sqrt{2} {(\Delta t)}^2
\right] 
\ \ . 
\label{eq:geon-corrgreen}
\end{equation}
As (\ref{eq:geon-corrgreen}) does not satisfy the Klein-Gordon
equation, we cannot interpret it as a correction in a free conformal
scalar field Green's function. Although the horizon-crossing geodesics
give the leading $a$-dependent correction in the 
Green's function, we therefore cannot use them to compute a
correction to the boundary stress-energy within our AdS/CFT model at least
for $d=5$.

\section{Discussion}
\label{sec:discussion}

In this paper we have analysed AdS/CFT correspondence for two
families of Einstein black holes in $d\ge4$ dimensions, modelling the
boundary CFT by a free conformal scalar field and evaluating the boundary
two-point function semiclassically from bulk geodesics. For the 
\hdbtz\ hole, which is locally 
AdS${}_d$ and generalises the nonrotating BTZ hole to $d\ge4$, the
model was fully self-consistent. The boundary state was the Euclidean
vacuum induced from the global vacuum on the conformal boundary of
AdS${}_d$, which is thermal in the sense appropriate for the
dS${}_{d-2}$ factor in the boundary metric. The boundary stress-energy
suggested a novel definition for the mass of the \hdbtz\ 
hole by AdS/CFT: In the
absence of an ADM mass, the interest in this definition
remains to be seen. We also analysed briefly a $\BbbZ_2$ quotient and a
generalisation involving rotation. 

For the generalised $d\ge4$ Schwarzschild-AdS hole with a flat $\BbbR^{d-2}$
horizon, the model was self-consistent at the
level of the boundary stress-energy, and the stress-energy had the 
thermal form in a temperature that agreed with the hole Hawking temperature
up to a $d$-dependent numerical factor. In particular, the energy density
was proportional to the hole ADM mass. However, the model
could not consistently accommodate corrections from a periodic horizon
dimension in the limit of large period with fixed mass for $d=5$. 
Similarly, the model could not consistently accommodate corrections from
geodesics that cross the horizons in a single-exterior version of
this $d=5$ hole, obtained as a $\BbbZ_2$ quotient. We suspect these
inconsistencies to be present for all $d\ge4$. 

The stress-energy tensor on the boundary of the \hdbtz\ hole received
contributions from bulk geodesics that pass through the near-horizon
region, and for the $\BbbZ_2$ quotient there were additional
contributions from geodesics that pass inside the horizon. These
`long' geodesics arise from the construction of the hole as a quotient
of AdS${}_d$ and have no counterpart in AdS${}_d$ itself. It follows
that these contributions represent an effect that is not present in a
boundary stress-energy tensor calculated from the near-infinity metric
by quasilocal techniques with counterterm subtraction
\cite{bala-kraus,kraus-larsen-siebelink,haro-skenderis-solo}: The
quasilocal stress-energy tensor is insensitive to quotients that
preserve the conformal rescaling near the infinity and cannot thus
depend on the parameter $\lambda_+$ that determines the size of the
hole. For the explicit computation of the quasilocal stress-energy for 
$d=5$, see~\cite{cai}. 
The parameter $\lambda_+$ would affect integration of the
quasilocal stress-energy, and this is how the quasilocal stress-energy
on the boundary of the BTZ hole reproduces the correct mass and
angular momentum as conserved charges~\cite{bala-kraus}, but the
absence of a global timelike Killing vector prevents a direct lift of
this BTZ result to the \hdbtz\ boundary. For the generalised
Schwarzschild-AdS hole, our stress-energy result for nonperiodic
horizon agrees with the quasilocal stress-energy~\cite{deh-khod}. 

Why did a period on the horizon of the generalised Schwarzschild-AdS 
hole make the model inconsistent? One might suspect the reason to have been
an inappropriate matching of the boundary and the bulk.\footnote{We thank
Rob Myers for raising this possibility.} 
In the semiclassical evaluation of the
boundary two-point function, the bulk geometry should be regarded as a 
saddle point in the gravitational path integral under boundary
conditions set by the boundary geometry. 
When the Euclidean-signature boundary has
two periodic dimensions, there are a priori two families of saddle points,
differing in the choice of the boundary circumference that is matched to the
bulk Euclidean time, and for the dominant saddle point this circumference
is the smaller
one~\cite{horo-myers,surya-schleich-witt,page-trans}.\footnote{For
similar issues with a finite distance boundary, 
see 
\cite{HartleWitt,LoukoRuback,GiuliniLouko}.} 
As we
considered a limit of large spatial period with fixed hole mass, our bulk
was in fact the dominant saddle point, so this cannot have caused the 
problem. 

The reason for the inconsistency may be just that a free conformal scalar
field is not a good boundary CFT model when bulk regions where the local 
geometry differs substantially from AdS${}_d$ become important: We see from 
(\ref{eq:per-smallrmin}) that in the limit of large period, the troublesome 
geodesics pass arbitrarily close to the horizon. 
For the ordinary 
Schwarzschild-AdS hole, it was indeed argued in
\cite{Fid-etal} that a better boundary CFT model is a high
conformal weight scalar field.\footnote{We thank Veronika Hubeny for
this observation.} 
By adjusting the parameter $\mu$
in~(\ref{eq:bulk-estimate}), our geodesic length analyses yield small
separation expansions for arbitrary conformal weight
Green's functions on the boundary, but gaining useful information from such
an expansion is more difficult when the field is not free. For a
discussion of the conformal anomaly, see~\cite{NO-anomaly}. 

It would be possible to evaluate boundary Wilson loops semiclassically for
our generalised Schwarzschild-AdS holes and analyse the
corrections that arise from the choice of the bulk topology. Our
Green's function results suggest, however, that one should first develop a
better understanding of what might constitute a reasonable boundary CFT
model for these bulks.

\section*{Acknowledgements}

We thank 
Simon Ross for discussions and the suggestion to look at the
\hdbtz{} holes, 
and 
Abhay Ashtekar, 
John Barrett, 
Ed Corrigan, 
Veronika Hubeny, 
Keijo Kajantie, 
Bernard Kay, 
Don Marolf, 
Rob Myers, 
Tony Sudbery 
and 
Reza Tavakol
for discussions and correspondence. 
Suggestions from an anonymous referee improved the manuscript. 
This work was
supported in part by EPSRC Fast Stream grant GR/R67170 and 
the University of Nottingham Research Committee.

\appendix

\section{Asymptotics for subsection \ref{subsec:hdbtz-geon}}
\label{app:b-asymptotics}

In this appendix we prove a lemma needed for the large $\abs{T}$
behaviour~(\ref{geonstress-largeT}). 

\begin{lemma}
\label{lemma:CtoE}
Let $\lambda_+>0$ and $p>0$. For $\alpha>0$, let 
\begin{subequations}
\label{eq:FG-defs}
\begin{align}
F(\alpha)
& :=
\sum_{k=0}^{\infty}
\frac{1}{{ \left\{ \cosh \bigl[(2k+1)\pi\lambda_{+} \bigr]
+\alpha \right\}}^{p}}
\ \ , 
\\
G(\alpha)
& :=
\sum_{k=0}^{\infty}
\frac{1}
{{\bigl[\frac{1}{2}{\rme}^{(2k+1)\pi\lambda_{+}}+\alpha\bigr]}^{p}}
\ \ . 
\end{align}
\end{subequations}
Then\/ 
$G(\alpha) = \mathcal{O}\left( \frac{\ln \alpha}{\alpha^p} \right)$
and\/ 
$F(\alpha) = G(\alpha) + \mathcal{O} \bigl(\alpha^{-(p+1)}\bigr)$
as
$\alpha\to\infty$. 
\end{lemma}

\myproof
Consider first~$G$. 
We write $G(\alpha) = {\bigl( 2 \rme^{-\pi\lambda_+}
\bigr)}^p S ( 2 \rme^{-\pi\lambda_+} \alpha )$, where 
\begin{equation}
S(x) := \sum_{k=0}^\infty 
\frac{1}{ {\bigl( \rme^{2 \pi \lambda_+ k} + x \bigr)}^p }
\ \ . 
\label{eq:S-def}
\end{equation}
Estimating the sum in (\ref{eq:S-def}) by the integral gives the
sandwich inequality 
$I(x) < S(x) < I(x) + {(1+x)}^{-p}$, 
where 
\begin{align}
I(x) 
&:= 
\int_0^\infty \frac{\rmd t}{{(\rme^{ 2 \pi \lambda_+ t} + x)}^p}
\nonumber
\\
& = 
\frac{1}{ 2 \pi \lambda_+ x^p} 
\int_{1/x}^\infty 
\frac{\rmd z}{z{(z+1)}^p}
\nonumber
\\
& = 
\frac{1}{ 2 \pi \lambda_+ x^p} 
\left\{
\int_{1/x}^\infty 
\frac{\rmd z}{z{(z+1)}}
\> 
+ \int_0^\infty 
\frac{\rmd z}{z} 
\left[ 
\frac{1}{{(z+1)}^p} - \frac{1}{{z+1}}
\right] 
\vphantom{\int_0^{1/x}}
\right. 
\nonumber
\\
& 
\ \ \ \ \ \ \ \ \ \ \ \ \ \ 
\left.
- 
\int_0^{1/x}
\frac{\rmd z}{z} 
\left[ 
\frac{1}{{(z+1)}^p} - \frac{1}{{z+1}}
\right] 
\right\}
\nonumber
\\
& = 
\frac{1}{ 2 \pi \lambda_+ x^p} 
\left\{
\ln x 
+ \int_0^\infty 
\frac{\rmd z}{z} 
\left[ 
\frac{1}{{(z+1)}^p} - \frac{1}{{z+1}}
\right] 
\> 
+ \mathcal{O}( x^{-1} )
\vphantom{\int_0^{\infty}}
\right\}
\ \ . 
\label{eq:z-lemma}
\end{align}
In (\ref{eq:z-lemma}) we have first changed variables by $\rme^{ 2 \pi
\lambda_+ t} = xz$ and then rearranged the integral in a form whose large
$x$ expansion is elementary. The sandwich inequality and
(\ref{eq:z-lemma}) imply 
$S(x) =
\mathcal{O}\left(
\frac{\ln x}{x^p} \right)$, and hence $G(\alpha) = \mathcal{O}\left(
\frac{\ln
\alpha}{\alpha^p} \right)$. 

Consider then the difference of $F$ and~$G$. 
From (\ref{eq:FG-defs}) we find 
\begin{equation}
\alpha^{p+1} \bigl[ G(\alpha)-F(\alpha) \bigr]
=
\sum_{k=0}^{\infty}
\alpha^{p+1} 
\left(
\frac{1}
{{\bigl[\frac{1}{2}{\rme}^{(2k+1)\pi\lambda_{+}}+\alpha\bigr]}^{p}}
- 
\frac{1}{{\left\{ \cosh \bigl[(2k+1)\pi\lambda_{+} \bigr]
+\alpha \right\}}^{p}}
\right)
\ \ , 
\label{eq:FG-diff}
\end{equation}
where the sum rearrangement is allowed by absolute convergence. An
elementary analysis shows that the $k$th term on the right-hand
side of (\ref{eq:FG-diff}) is positive and bounded above by
$\frac{1}{2}p{\rme}^{-(2k+1)\pi\lambda_{+}}$ and tends to
$\frac{1}{2}p{\rme}^{-(2k+1)\pi\lambda_{+}}$ as $\alpha\to\infty$. It
follows by dominated convergence that we can take the limit
$\alpha\to\infty$ in (\ref{eq:FG-diff})
termwise, and summing the geometric series gives
\begin{equation}
\alpha^{p+1} \bigl[ G(\alpha)-F(\alpha) \bigr]
\xrightarrow[\alpha\to\infty]{}
\frac{p}{4\cosh(\pi\lambda_{+})}
\ \ . 
\end{equation}
Hence $F(\alpha) = 
G(\alpha) + \mathcal{O} \bigl(\alpha^{-(p+1)}\bigr)$. 
$\blacksquare$

\section{Expansion of $\Lren$}
\label{app:L-expansion}

In this appendix we solve the system (\ref{eq:Ltx-ren}) for $\Lren$ to
next-to-leading order in the limit of small 
$\Delta\tau$ and $\Delta\vec{x}$ with fixed 
$\abs{\Delta\vec{x}}/\abs{\Delta\tau}$. 

We specify the geodesic in (\ref{eq:Ltx-ren}) by $\vec{C}$
and~$\rmin$. From~(\ref{poly}), 
$C$~is then given by 
\begin{equation}
C \myell = \sqrt{ 
\bigl( \rmin ^{2}-\vec{C}^{2} \bigr) 
\bigl[1- (r_{h}/\rmin)^{d-1} \bigr]
} 
\label{eq:C-function}
\ \ , 
\end{equation}
and $P(r)$ factorises as 
\begin{subequations}
\label{eq:P-factorisation}
\begin{equation}
P(r)=(r-\rmin )(r^{d}+a_{d-1}r^{d-1}+\cdots+a_{0})
\ \ , 
\end{equation}    
where 
\begin{align}
a_{d-1} 
&=
\rmin 
\ \ , 
\nonumber 
\\
a_{d-i}
&=
\frac{r_{h}^{d-1}}{\rmin ^{d-i-1}}
\left( 1-\frac{\vec{C}^{2}}{\rmin^{2}} \right)
\ \ , \ \ 2 \le i \le d-2 \ \ , 
\nonumber 
\\
a_{1}
&=
-\frac{r_{h}^{d-1}\vec{C}^{2}}{\rmin ^{2}}
\ \ , 
\nonumber 
\\
a_{0}
&=
-\frac{r_{h}^{d-1}\vec{C}^{2}}{\rmin }
\ \ . 
\end{align}
\end{subequations}

We first expand 
(\ref{dt-ren}) 
and (\ref{dx-ren}) to next-to-leading order at 
large~$\rmin$, keeping $\vec{C}/\rmin$
constant. From (\ref{dx-ren}), we obtain 
\begin{align}
\Delta \vec{x} 
&= 
2\vec{C}\myell^2 \int^{\infty}_{\rmin }
\frac{r^{\frac{d-5}{2}}
\rmd r}{\sqrt{(r-\rmin )(a_{0}+\cdots+r^{d})}} 
\nonumber
\\  
&= 
2\vec{C} \myell^2 \int_{\rmin
}^{\infty} \frac{\rmd r}{r^{2}\sqrt{r^{2}-\rmin
^{2}}
\sqrt{ 1+\frac{a_{d-2}}{r(r+\rmin )}+ 
\cdots+\frac{a_{0}}{r^{d-1}(r+\rmin )}
}
} 
\nonumber
\\
&\sim 
2\vec{C}\myell^2 \left[ \int_{\rmin }^{\infty} \frac{\rmd
r}{r^{2}\sqrt{r^{2}-\rmin ^{2}}} - 
\sum_{i=0}^{d-2} \frac{a_{i}}{2} \int_{\rmin }^{\infty} 
\frac{\rmd r}{r^{d+1-i}(r+\rmin )\sqrt{r^{2}-\rmin ^{2}}} \right]
\ \ . 
\label{app.x}
\end{align}
Note that all the terms under the sum in (\ref{app.x}) contribute to the
next-to-leading term, which is suppressed compared to the leading term by
the factor 
${(r_{h}/\rmin)}^{d-1}$. The dropped terms are suppressed compared
with the leading term by ${(r_{h}/\rmin)}^{2(d-1)}$. 
A~similar treatment of (\ref{dt-ren}) yields 
\begin{equation}
\Delta \tau 
\sim 
2C\myell^{3} \int_{\rmin }^{\infty} 
\frac{\left[1+\frac{r_{h}^{d-1}}{r^{d-1}}-
\sum_{i=0}^{d-2}\frac{a_{i}}{2r^{d-1-i}(r+\rmin
)}\right] \rmd r}{r^{2}\sqrt{r^{2}-
\rmin ^{2}}}
\ \ , 
\label{app.t}
\end{equation} 
where $C$ can be replaced by its next-to-leading order expansion 
\begin{equation}
C\myell \sim 
\sqrt{\rmin ^{2}-\vec{C}^{2}} \left(1-\frac{r_{h}^{d-1}}{2\rmin
^{d-1}}\right)
\ \ .
\end{equation}
The integrals in (\ref{app.x}) and (\ref{app.t}) can be evaluated by the
identities, easily proved by induction, 
\begin{subequations}
\label{ids12} 
\begin{gather}
\int_{\rmin }^{\infty} 
\frac{\rmd r}{r^{m+1}\sqrt{r^{2}-\rmin ^{2}}} 
= 
\frac{1}{\rmin ^{m+1}} 
\frac{(m-1)!!}{m!!} 
\left(\frac{\pi}{2}\right)^{\frac{1}{2}[1+(-1)^{m}]}
\ \ , 
\\
\int_{\rmin }^{\infty} 
\frac{\rmd r}{r^{m}(r+\rmin )\sqrt{r^{2}-\rmin ^{2}}} = 
\frac{1}{\rmin ^{m+1}}
\sum_{i=0}^{m}
(-1)^{i} \frac{(m-i-2)!!}{(m-i-1)!!} 
\left(\frac{\pi}{2}\right)^{\frac{1}{2}[1-(-1)^{m-i}]}
\ \ , 
\end{gather}
\end{subequations}
where $m \geq 0$ and the double factorial of a negative number is
understood to be unity. 
Collecting, we obtain 
\begin{subequations}
\label{eq:Dtx} 
\begin{align}
\Delta \tau  
& \sim 
\frac{2\myell^{2}}{\rmin ^{2}} 
\sqrt{\rmin ^{2}-\vec{C}^{2}}
\nonumber 
\\
& 
\ \ \ \ 
\times 
\left[ 1 + \frac{\alpha_d}{2} 
\! 
\left( \frac{\vec{C}^{2}}
{\rmin ^{2}}
\right) 
\frac{r_{h}^{d-1}}{\rmin ^{d-1}} +
\bigl(\alpha_d -\tfrac{1}{2}\bigr) 
\frac{r_{h}^{d-1}}{\rmin ^{d-1}} 
- \frac{\beta_d}{2} \! \left( 1-\frac{\vec{C}^{2}}{\rmin
^{2}} \right)
\frac{r_{h}^{d-1}}{\rmin ^{d-1}} \right] 
\ \ , 
\label{Dt} 
\\
\Delta \vec{x} 
& \sim 
\frac{2\vec{C}\myell^2}{\rmin ^{2}} 
\left[ 1 + \frac{\alpha_d}{2} 
\! 
\left( 
\frac{\vec{C}^{2}}{\rmin ^{2}}
\right) 
\frac{r_{h}^{d-1}}{\rmin ^{d-1}} - 
\frac{\beta_d}{2} \! \left(1-\frac{\vec{C}^{2}}{\rmin ^{2}} \right)
\frac{r_{h}^{d-1}}{\rmin ^{d-1}} \right]
\ \ , 
\label{Dx}
\end{align}
\end{subequations}
where 
\begin{subequations}
\begin{align}
\alpha_d 
&:=
\frac{(d-1)!!}{d!!} 
\left(\frac{\pi}{2}\right)^{\frac{1}{2}[1+(-1)^{d}]}
\ \ , 
\\
\beta_d 
&:= 
[1+(-1)^{d}] \left( \frac{3\pi}{8}-1 \right)
+ 
\sum_{k=1}^{[\frac{d-3}{2}]}\frac{(d-2k-1)!!}{(d-2k)!!} 
\left(\frac{\pi}{2}\right)^{\frac{1}{2}[1+(-1)^{d}]} 
\ \ . 
\label{eq:beta-d-def}
\end{align}
\end{subequations}
The square  bracket in the sum limit in (\ref{eq:beta-d-def})
denotes integer part. 

Next, we invert (\ref{eq:Dtx}) for $\vec{C}$ and $\rmin$ to the
next-to-leading order in the limit of small $\Delta\tau$ and
$\Delta\vec{x}$ with fixed $\abs{\Delta\vec{x}}/\abs{\Delta\tau}$. 
The result is 
\begin{subequations}
\label{eq:rmin-Cvec-sol} 
\begin{align}
\rmin
&\sim
\frac{2\myell^{2}}{D}
\left\{
1+
\frac{MD^{d-3}}{2^{d-1}\myell^{2d-4}}
\bigl[
\alpha_d (\Delta\vec{x})^{2} 
+
\bigl(2\alpha_d-\beta_d-1\bigr)
(\Delta \tau)^{2} 
\bigr]
\right\}
\ \ , 
\label{rmin} 
\\
\frac{\vec{C}^{2}}{\rmin^{2}}
&\sim
\frac{(\Delta\vec{x})^{2}}{D^{2}}
\left[
1+
\frac{MD^{d-3}}{2^{d-2}\myell^{2d-4}}
\bigl( 2\alpha_d - 1 \bigr) (\Delta \tau)^{2} 
\right]
\ \ ,  
\label{vecC}
\end{align}
\end{subequations}
where $D := \sqrt{(\Delta \tau)^{2} + (\Delta \vec{x})^{2}}$. 
Comparison of (\ref{eq:rmin-Cvec-sol}) and our large $\rmin$ expansion of 
(\ref{dt-ren}) 
and (\ref{dx-ren}) shows that this large
$\rmin$ expansion kept all terms that contribute to the next-to-leading
order in $\Delta\tau$ and
$\Delta\vec{x}$. Hence (\ref{eq:rmin-Cvec-sol}) solves (\ref{dt-ren}) 
and (\ref{dx-ren}) to the order shown. 

Finally, the integral in (\ref{eq:Lren-def}) can be expanded by similar
techniques. We omit the details. Using (\ref{eq:rmin-Cvec-sol}) and
writing the remaining double factorial in terms of the gamma-function 
gives~(\ref{length}).

\section{Finite temperature stress-energy with one 
periodic dimension}
\label{perMink}

In this appendix we find the leading finite size correction to the
finite temperature stress-energy tensor of a free conformal scalar
field on Minkowski spacetime with one periodic spacelike
dimension. Despite the interest of finite size effects in finite
temperature field theory (see for example~\cite{hasen-leut}), we have
found this correction in the literature only in dimension two
(\cite{birrell}, Section~4.2). In four dimensions, the electromagnetic
finite temperature Casimir effect between two perfectly conducting planes
is closely similar~\cite{brown-maclay}. 


Following the conventions of the main text, 
the Minkowski dimension is $d-1$ with $d>3$. 
The coordinates on the
Euclidean-signature section are 
$(\tau,x,\vec{y})$ with $\vec{y} \in \BbbR^{d-3}$ and 
$(\tau,x,\vec{y}) \sim (\tau,x+a,\vec{y})$. 
The inverse temperature is~$\beta$. 

Let 
\begin{equation}
\Gwide_0 (\tau,x,\vec{y})
:= 
\frac{1}{{\bigl({\tau}^{2}+{x}^{2}
+{\vec{y}}^{2}\bigr)}^{(d-3)/2}}
\ \ . 
\end{equation}
By the method of images, the Green's function is 
\begin{equation}
G = 
\frac{\Gamma \bigl(\frac{d-3}{2} \bigr) }
{4{\pi}^{(d-1)/2} } \Gwide
\ \ , 
\end{equation}
where 
\begin{align}
\Gwide (\tau,x,\vec{y})
&= 
\Gwide_0 (\tau,x,\vec{y})
\nonumber 
\\
&
\ + 
\sideset{}{'}\sum_{m,n}
\Bigl[ 
\tfrac12 \Gwide_0 (\tau+m\beta , x+na , \vec{y})
+ 
\tfrac12 \Gwide_0 (\tau-m\beta , x-na , \vec{y})
- 
\Gwide_0 (m\beta , na , \vec{0}) 
\Bigr] 
\label{eq:Gtilde-def}
\end{align}
and the primed sum is over $(m,n) \in \BbbZ^2
\setminus \{(0,0)\}$. 
The inclusion of the constant terms and the
$(m,n)\leftrightarrow(-m,-n)$ pairing 
make the sum convergent in
absolute value for all $d>3$.

\begin{lemma}
\label{lemma:g-asymptotics}
Let $p>1$. For $u>0$, let 
\begin{equation}
g_{p}(u) :=
\sum_{m,n=1}^{\infty}\frac{1}{{(m^{2}+n^{2}u^2)}^{p}}
\ \ . 
\label{eq:gq-def}
\end{equation}
Then 
\begin{equation}
g_{p}(u)
= 
\frac{ \sqrt{\pi} \, \Gamma \bigl(p-\frac{1}{2}\bigr) \zeta(2p-1)}
{2\Gamma(p) \, u^{2p-1}}
+ \mathcal{O} \bigl( u^{-2p} \bigr) 
\label{eq:gqsum-as}
\end{equation}
as\/ $u\to\infty$, where $\zeta$ is the Riemann zeta-function, and the
expansion (\ref{eq:gqsum-as}) can be differentiated in~$u$. 
\end{lemma}

\myproof
An integral estimate similar to that in the proof of Lemma
\ref{lemma:CtoE} yields 
\begin{equation}
\sum_{m=1}^{\infty}\frac{1}{{(m^{2}+v^2)}^{p}}
= 
\frac{ \sqrt{\pi} \, \Gamma \bigl(p-\frac{1}{2}\bigr)}
{2\Gamma(p) \, v^{2p-1}} 
+ \mathcal{O} \bigl( v^{-2p} \bigr) 
\label{eq:gqsum-lemma}
\end{equation}
as $v\to\infty$, where the 
coefficient in the leading term arises as the integral 
$\int_{0}^{\infty}
{(1+y^{2})}^{-p} \rmd y$~\cite{gradstein}. 
Setting in (\ref{eq:gqsum-lemma}) $v=nu$ and summing over~$n$,
(\ref{eq:gqsum-as}) follows using the infinite sum representation of the
zeta-function~\cite{gradstein}. 
To justify differentiation of~(\ref{eq:gqsum-as}), multiply
(\ref{eq:gqsum-lemma}) by powers of~$n^2$, 
set $v=nu$ and sum over~$n$, and compare
with derivatives of~(\ref{eq:gq-def}). 
$\blacksquare$


We split $\Gwide$ as 
\begin{equation}
\Gwide = \Gwide_{0} + \Gwide_{1} + \Gwide_{2} + \Gwide_{3}
\ \ ,
\end{equation}
where $\Gwide_{1}$ consists of
the $n=0\ne m$ terms, $\Gwide_{2}$ consists of the $m=0\ne n$ terms 
and
$\Gwide_{3}$ consists of the $m\ne0\ne n$ terms. 
In $\Gwide_{1}$, $\Gwide_{2}$ and $\Gwide_{3}$ we expand the summands to
quadratic order in $\tau$, $x$ and~$\vec{y}$. 
Interchanging the sum and the
expansion, justified by convergence arguments that we omit here, 
we find 
\begin{subequations}
\begin{align}
\Gwide_{1}^{(2)}
&=
\frac{(d-3)\zeta(d-1)}
{\beta^{d-1}}\left[(d-2)\tau^{2}-x^{2}-\vec{y}^{2}\right]
\ \ , 
\\
\Gwide_{2}^{(2)}
&= 
\frac{(d-3)\zeta(d-1)}{a^{d-1}}\left[(d-2)x^{2}
-\tau^{2}-\vec{y}^{2}\right]
\ \ , 
\\
\Gwide_{3}^{(2)} 
&=
2(d-3) 
\sum_{m,n=1}^{\infty}
\left[\frac{(d-2) \tau^{2}-x^{2}-
\vec{y}^{2}}
{{(m^{2}\beta^{2}+n^{2}a^{2})}^{(d-1)/2}}
+\frac{(d-1)n^{2}a^{2} \bigl( x^{2}-\tau^{2} \bigr)}
{{(m^{2}\beta^{2}+n^{2}
a^{2})}^{(d+1)/2}}\right]
\nonumber
\\
&=
2(d-3) 
\beta^{-(d-1)} 
\Bigl\{
\bigl[ (d-2)\tau^{2}-x^{2}-\vec{y}^{2}
\bigr]
g_{(d-1)/2} \bigl( a \beta^{-1} \bigr)
\nonumber 
\\
&
\hspace{20ex}
+ 
\bigl( \tau^{2} - x^{2} \bigr) 
a \beta^{-1} 
g_{(d-1)/2}' 
\bigl( a \beta^{-1} \bigr)
\Bigr\}
\ \ , 
\label{eq:Gwide3}
\end{align}
\end{subequations}
where the superscript ${}^{(2)}$ indicates that only terms up to quadratic
order have been kept and the prime on $g_{(d-1)/2}$ denotes derivative
with respect to the argument. 

In the limit of large $a$ with fixed~$\beta$,
(\ref{eq:gqsum-as}) and (\ref{eq:Gwide3}) imply 
\begin{equation}
\Gwide_{3}^{(2)}
=
\frac{ 2 \sqrt{\pi} \, \Gamma \bigl(\frac{d-2}{2}\bigr) \zeta(d-2)}
{ \Gamma \bigl( \frac{d-3}{2} \bigr) \beta a^{d-2} }
\left[ (d-3)x^{2}-\vec{y}^{2} \right] 
+\mathcal{O} \bigl( a^{-(d-1)} \bigr)
\ \ . 
\end{equation}
Continuing $G$ to Lorentz-signature,  standard point-splitting 
methods \cite{birrell} give 
\begin{align}
\langle T_{\mu\nu} \rangle 
&= 
\frac{ \Gamma \bigl( \frac{d-1}{2} \bigr) \zeta(d-1) }
{ {\pi}^{(d-1)/2} \beta^{d-1}} 
\times 
{\mathrm{diag}} ( d-2,1,\ldots,1 )
\nonumber
\\
&
\ \ \ \ \ 
+ 
\frac{ \Gamma \bigl( \frac{d-2}{2} \bigr) \zeta(d-2) }
{ {\pi}^{(d-2)/2} \beta a^{d-2}} 
\times 
{\mathrm{diag}} ( 0, 3-d ,1,\ldots,1 ) 
\nonumber
\\
&
\ \ \ \ \ 
+\mathcal{O} \bigl( a^{-(d-1)} \bigr)
\ \ , 
\label{eq:perMink-T-leading}
\end{align}
where the first term is the usual Minkowski value, coming from
$\Gwide_{1}^{(2)}$, and the second term is the leading correction, coming
from~$\Gwide_{3}^{(2)}$. Note that this correction dominates the 
zero-temperature vacuum polarisation term that comes from
$\Gwide_{2}^{(2)}$ and is proportional to~$a^{-(d-1)}$.

\section{Kruskal extension}
\label{app:kruskal}

In this appendix we present the Kruskal-type extension of the
metric~(\ref{eq:gwmetric}). 
For notational convenience we
write $d= n+1$, $n\ge3$. 

Starting in the exterior region $r>r_h$, we define the tortoise
coordinate $r_{*}$ by 
\begin{align}
r_{*} 
&:= 
- \int_r^{\infty} \frac{\rmd \tilde{r}}{f({\tilde{r})}}
\nonumber 
\\
&= 
\frac{\myell^2}{n r_h} 
\left[ 
\ln \! \left( \frac{r - r_h}{r + r_h} \right)
- h(r_h/r)
\right]  
\ \ , 
\end{align}
where 
\begin{equation}
h(s) := \int_0^s 
\left( 
\frac{n}{1-z^n} 
- 
\frac{2}{1-z^2} 
\right) 
\rmd z
\ \ , 
\label{eq:h-def}
\end{equation}
and the Kruskal coordinates $(U,V,\vec{x})$ by 
\begin{subequations}
\begin{align}
U
&:=
-\exp\left[-\frac{n r_{h}(t-r_{*})}{2\myell^{2}}\right]
\ \ , 
\\
V
&:=
\exp\left[\frac{n r_{h}(t+r_{*})}{2\myell^{2}}\right]
\ \ . 
\end{align}
\end{subequations} 
The metric reads 
\begin{subequations}
\label{eq:kruskalmetric}
\begin{equation}
\rmd s^{2} 
= 
- 
\frac{4\myell^{2} (r+r_h) \bigl( r^n - r_h^n \bigr)}
{n^2 r_h^2 r^{n-2} (r-r_h)} 
\rme^{h(r_h/r)} 
\, 
\rmd U \rmd V 
+ 
\frac{r^{2}}{\myell^2} 
\, \rmd\vec{x}^{2}
\ \ , 
\end{equation}
where $r$ is determined as a function of $U$ and $V$ by 
\begin{equation}
-UV = 
\left( \frac{r - r_h}{r + r_h} \right)
\rme^{-h(r_h/r)} 
\ \ . 
\end{equation}
\end{subequations}
The function $h(s)$ (\ref{eq:h-def}) is nonsingular for $0 \le s<\infty$, 
$h(0)=0$ and $\lim_{s\to\infty}h(s) = \pi \cot(\pi/n)$, 
using 3.241.3 in~\cite{gradstein}.\footnote{For an
explicit expression of $h(s)$ in terms of elementary functions, see
2.144 in~\cite{gradstein}.} It follows by 
standard
considerations that 
(\ref{eq:kruskalmetric}) is a global chart with $-1 < UV <
\rme^{-\pi \cot(\pi/n)}$. 

As the infinities are at $UV \to -1$ and the singularities are at $UV \to
\rme^{-\pi \cot(\pi/n)} < 1$, 
the singularities in the conformal diagram cave inward relative to the
infinities~\cite{strobl}. In the AdS/CFT context, this gives rise to
phenomena analysed for Schwarzschild-AdS holes in~\cite{Fid-etal}.

\end{document}